\documentclass[twocolumn,prb,showpacs,superscriptaddress,preprintnumbers,amsmath,amssymb]{revtex4}
\usepackage{graphicx}
\usepackage{mathptmx}
\usepackage{verbatim}
\usepackage{epstopdf}
\usepackage{amsmath}
\usepackage{amssymb}
\usepackage{mathdots}
\usepackage{amsfonts}
\usepackage{mathrsfs}
\usepackage{color}
\usepackage{bm} 
\usepackage{xspace}
\usepackage{dcolumn}

\begin{document}
\title{
Electronic and magnetic properties of single-layer MPX$_3$ metal phosphorous trichalcogenides
}
\author{Bheema Lingam Chittari}
\affiliation{SKKU Advanced Institute of Nanotechnology, Sungkyunkwan University, Suwon, 16419, Korea}
\affiliation{Department of Physics, University of Seoul, Seoul 02504, Korea}
\author{Youngju Park}
\affiliation{Department of Physics, University of Seoul, Seoul 02504, Korea}
\author{Dongkyu Lee}
\affiliation{Department of Physics, University of Seoul, Seoul 02504, Korea}
\author{Moonsup Han}
\affiliation{Department of Physics, University of Seoul, Seoul 02504, Korea}
\author{Allan H. MacDonald}
\affiliation{Department of Physics, The University of Texas at Austin, Austin, Texas 78712, USA }
\author{Euyheon Hwang}
\email{euyheon@skku.edu}
\affiliation{SKKU Advanced Institute of Nanotechnology, Sungkyunkwan University, Suwon, 16419, Korea}
\author{Jeil Jung}
\email{jeiljung@uos.ac.kr}
\affiliation{Department of Physics, University of Seoul, Seoul 02504, Korea}

\begin{abstract}
We survey the electronic structure and magnetic properties of two 
dimensional (2D) MPX$_3$ (M= V, Cr, Mn, Fe, Co, Ni, Cu, Zn, and X = S, Se, Te)
transition metal chalcogenophosphates to shed light on their potential role 
as single-layer van der Waals materials that possess magnetic order. 
Our {\em ab initio} calculations predict that most of these single-layer materials are antiferromagnetic semiconductors.  
The band gaps of the antiferromagnetic states decrease as the 
atomic number of the chalcogen atom increases (from S to Se to Te),
leading in some cases to half-metallic ferromagnetic states or to non-magnetic metallic states. 
We find that the competition between antiferromagnetic and ferromagnetic
states can be substantially influenced by gating and by strain engineering.
The sensitive interdependence we find between magnetic, structural, and electronic properties
establishes the potential of this 2D materials class for applications in spintronics.
\end{abstract}


\maketitle
\section{Introduction}
Following seminal studies that reported on the exfoliation of a stable single-layer graphene and on 
transport measurements with nearly ideal Dirac fermion fingerprints,\cite{novoselov_nature,philipkim_nature}
research on ultrathin two dimensional (2D) materials has emerged during the last decade as one of the most 
active research topics in condensed matter physics. 
Two-dimensional materials are interesting in part because their properties can be modified {\it in situ} by adjusting gate voltages. 
Recently, the focus of 2D materials research
has expanded beyond graphene to include other layered van der Waals materials with 
a variety of distinct physical properties.~\cite{novoselov_pnas}
For example, one popular class of 2D materials is the transition metal dichalcogenides (TMDC) \cite{chowalla}
which includes metals, semiconductors with exceptionally strong 
light-matter coupling,~\cite{lightmatter} and broken symmetry electronic states 
including ones with charge density wave and superconducting order.~\cite{tas2,tmdneto,frindt,mos2sc}
Magnetic order has, however, not yet been established in any two-dimensional material.  
Room-temperature magnetism in a single-layer material is an extremely attractive 
materials target because of the expectation that it might provide unprecedented electrical 
control of magnetism and enable new classes of information processing devices that
incorporate non-volatile memory elements more intimately. 

One strategy to search theoretically for magnetism in isolated two-dimensional van der Waals materials is to explore the magnetic properties
of single-layers exfoliated from a bulk material that exhibits robust magnetic order. 
Following this approach, recent theoretical studies have proposed a number of 
potential magnetic single-layer van der Waals materials, 
including group-V based dichalcogenides,~\cite{2dmagtmd}
FeBr$_3$, the chromium based ternary 
tritellurides CrSiTe$_3$ and CrGeTe$_3$,~\cite{lebegue, max1, max2, max3, max4, max5, max6,sivadas}
CrX$_3$ trihalides,~\cite{cri3,crx3} and MnPX$_3$  ternary chalcogenides.~\cite{niu,jacs}
{The vanadium based dichalcogenides, VX$_2$ (X=S, Se) were proposed first 
and are predicted to have strain tunable ferromagnetic phases.~\cite{2dmagtmd}
The trihalides CrX$_3$ (X = F, Cl, Br, I)~\cite{crx3} are new classes of
semiconducting ferromagnet with Curie temperature predictions of T$_{\rm C}$ $<$ 100~K.

The bulk ternary tritellurides CrATe$_3$ (A = Si, Ge)~\cite{max1} were predicted
within LDA to be ferromagnetic with small band gaps of 0.04 and 0.06 eV respectively.  
Their few layers limits have recently been studied in temperature dependent transport experiments.~\cite{max6}
The anisotropy along the c-axis and the dynamic correlations in the ab-plane seen by elastic 
and inelastic neutron scattering are  characteristic of 2D magnetism.~\cite{max3} 
In the single layer limit CrSiTe$_3$ is a semiconductor with a GGA gap of 0.4~eV,~\cite{max5} substantially larger than its bulk value,
and has a negative thermal expansion.~\cite{max2} 
Reports differ on the most stable magnetic phase 
between ferromagnetic~\cite{max1,max2,max5} or antiferromagnetic.~\cite{sivadas}
For larger atomic number compounds like  
CrGeTe$_3$ and CrSnTe$_3$,  DFT predicts ferromagnetic semiconducting phases
with Curie temperatures between 80-170K.~\cite{sivadas,max4,max6}}

{We focus here on single-layer materials formed from 
compounds in the transition metal phosphorous trichalcogenide (MPX$_3$) family,
which are known to exhibit magnetism for M=Mn and for a number of other metal atom species.  
Metal phosphorous trichalcogenides are cousins of CrSiTe$_3$, but are so far less studied for 2D magnetism. 
According to one recent study monolayer MnPSe$_3$ and MnPS$_3$ exhibit Neel antiferromagnetism~\cite{jacs} 
and valley-dependent optical properties.~\cite{niu}}
The structures and  magnetic properties of
some bulk compounds from this family have already been extensively studied.~\cite{mpx3struct1,wildes,mpx3struct2,mpx3struct3,mpx3struct4,mpx3struct5,
mpx3mag1,mpx3mag2,mpx3mag3,mpx3mag4,mpx3mag5,mpx3mag6,mpx3mag7,mpx3mag8,mpx3mag9,
mpx3mag10,mpx3mag11,mpx3mag12,mpx3mag13,mpx3mag14,mpx3other15,mpx3other16,mpx3other16b,mpx3other16c,mpx3other17,mpx3other18}
Because of the van der Waals character of these materials, one focus of bulk 
materials research is to characterize ion intercalation properties.~\cite{mpx3inter1,mpx3inter2,mpx3inter3,mpx3inter4,mpx3inter5}
For 2D magnetic materials to be most useful in device applications it is desirable to seek pathways to increase the critical temperature
at which they order magnetically.
We compare theoretical predictions for a variety of
late 3d transition metals (M= V, Cr, Mn, Fe, Co, Ni, Cu, Zn) 
and consider all there chalcogen atoms (X = S, Se, Te) in an effort to explore
the magnetic phases that can be expected as fully as possible.  
We study how the electronic bands are modified when the magnetic state undergoes 
a transition from antiferromagnetic to ferromagnetic, or from magnetic to non-magnetic.
Our results confirm the expected strong interdependence between magnetism and structural properties, 
for example lattice constant and crystal symmetry, and explain a surprisingly strong dependence of exchange interaction 
strengths on electron density and strain.  Because these materials may have relatively strong 
correlations, {\it ab initio} density-functional theory is not able to make quantitatively reliable 
predictions for all properties.  Nevertheless our survey provides considerable insight into 
materials-property trends, and into the potential for engineering the magnetic properties of these 
materials using field effects and strain engineering.  

Our paper is structured in the following manner.
We start in Sec.~\ref{sec:methods} by briefly summarizing some specific details 
of our first principles electronic structure calculations.
In Sec.~\ref{sec:groundstate} we discuss our results for ground-state properties
including structure, magnetic properties, and electronic band structures and densities-of-states. 
Sec.~\ref{sec:tunability} is devoted to an analysis of the 
carrier-density dependence of the magnetic ground state,
and to a study of the influence of strain on the magnetic phase diagram. 
Finally we close the paper in Sec.\ref{sec:summary} with a summary and discussion
of our results.

\section{{\em Ab initio} calculation details}
\label{sec:methods}
The study of the ground-state electronic structure and magnetic properties in this work 
has been carried out using plane-wave density functional theory as implemented in Quantum Espresso.~\cite{espresso}
We have used the Rappe-Rabe-Kaxiras-Joannopoulos ultrasoft (RRKJUS) pseudoptentials
for the semi-local Perdew-Burke-Ernzerhof (PBE) generalized gradient approximation (GGA)~\cite{GGA} together with the VdW-D2 
correction~\cite{d2grimme}.  We choose the GGA+D2 as a reference calculation because of the  
overall improvement of the GGA over the LDA~\cite{LDA} for
covalent bond description, and add the longer ranged D2 correction to improve the 
description of binding between the layers. 
(The GGA typically performs poorly for van der Waals bonds.~\cite{noamarom})
The magnetic solutions have also been compared with calculations employing the DFT+U scheme,
using the same value of $U = 4$~eV and alternatively 
using onsite repulsion $U$ values that saturate the magnitude of the band gap. 
Comparisons were made with hybrid HSE+D2 functionals~\cite{hse}
to assess non-local exchange effects that could further influence the magnetic ground states. 
All structures were optimized without constraints until the forces on each atom reached 10$^{-5}$~Ry/au.
The self-consistency criteria for total energies has been set at 10$^{-10}$~Ry and 
momentum space integrals were performed using a regularly spaced
k-point sampling density of 16$\times$16$\times$1 for the triangular lattice case 
and 16$\times$8$\times$1 for the rectangular lattice case, 
with a plane wave energy cutoff between 60 to 90~Ry. 
For the HSE+D2 calculation we used a coarser effective k-point sampling of 4$\times$4$\times$1. 
The out-of-plane vertical size of the periodic supercell was chosen to be 25~$\AA$,
which typically leaves an adequate vacuum spacing greater than 10~$\AA$ between
two-dimensional layers.

\begin{figure}[htb!]
\begin{center}
\includegraphics[width=8cm]{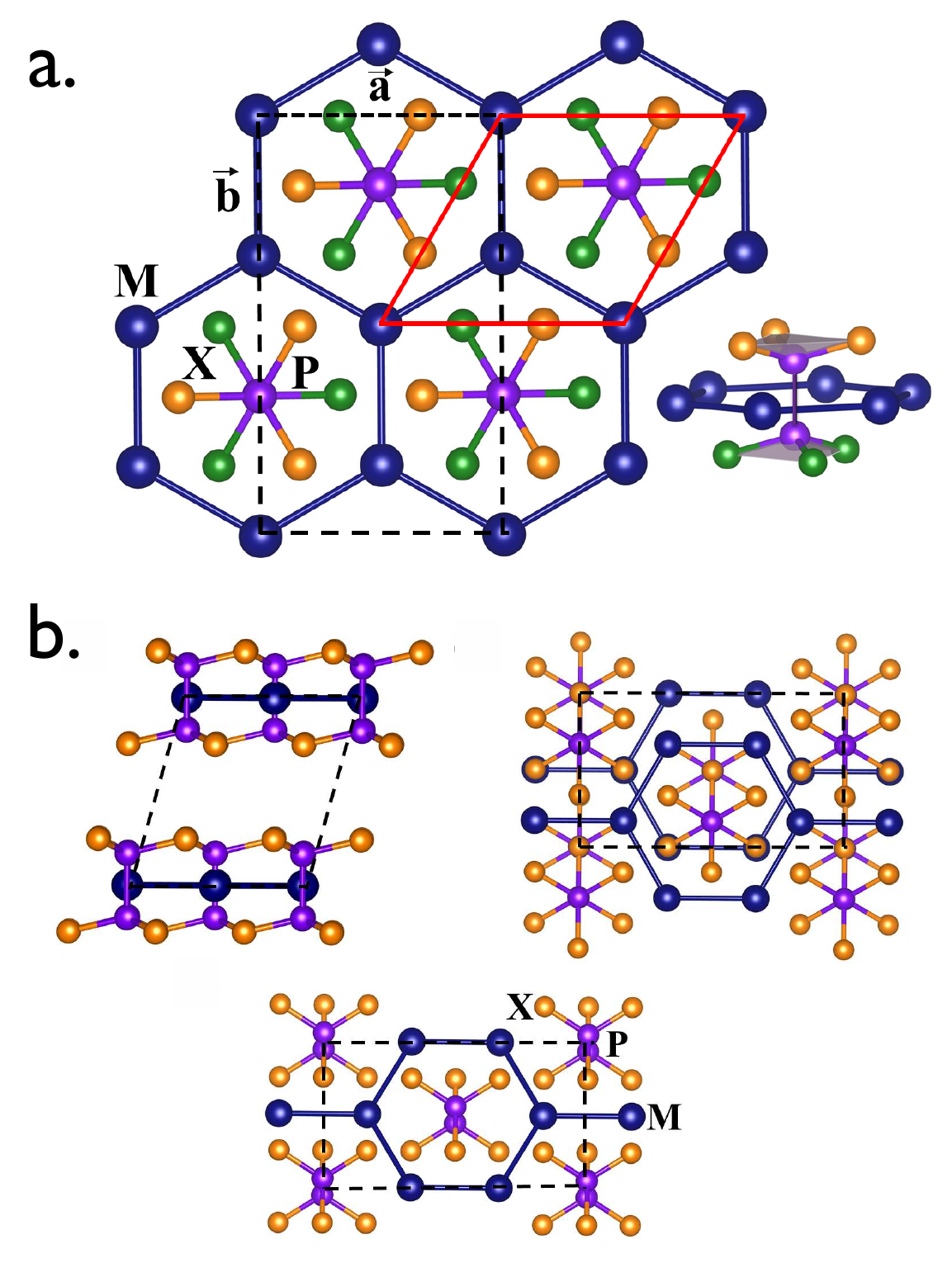}
\caption{(Color online) 
Schematic structure of the transition metal phosphorous trichalcogenide MPX$_3$ compounds with
(P$_2$X$_6$)$^{4-}$ bipyramids that enclose metal atoms. 
{\bf a.} Atomic structure of a MPX$_3$ monolayer. 
The rectangular supercell referred to in the main text
is indicated by black dashed lines and the smaller triangular lattice unit cell is 
represented by red solid lines.  Each triangular unit cell contains a transition metal atoms(M) 
honeycomb sublattice.
The phosphorus (P) dimers are perpendicular to the plane at the center of each 
honeycomb lattice hexagon, and three sulphur atoms are bound to each of the P atoms.
The top and bottom chalcogen trimers have a relative in-plane twist of 60$^\circ$ degrees.
{\bf b.} 
The lowest energy 3D-bulk structure consist of individual layers stacked in an ABC sequence.
We present the structure from three viewpoints, a side view, a top in which the phosphorous atoms 
within a single layer lie on top of each other, and a view in which
the transition metal atoms of adjacent layers lie 
on top of each other.}\label{fig:cellstructure}    
\end{center}
\end{figure}

\section{Structural and magnetic properties}
\label{sec:groundstate}
The atomic structure of MPX$_3$ transition metal chalcogenophosphate layers is anchored by 
(P$_2$X$_6$)$^{4-}$ bipyramids arranged in a triangular lattice that provide enclosures for transition metal atoms. 
(See Fig.~\ref{fig:cellstructure} for a schematic illustration of the single layer unit cell and 
the bulk atomic structure.) 
The bulk crystals consists of ABC-stacked single layer assemblies that are held together by van der Waals forces.
Although the atomic structures of single layer transition metal phosphorous trichalcogenide crystals are similar to those of bulk crystals,
small changes appear in response to the absence of the interlayer coupling,
with distortions in the ground-state crystal geometries correlated mainly with
the magnetic phase.
The analysis of magnetic properties is simplified by the fact that the magnetic moments develop almost 
entirely at the metal atom sites. 

We calculate the magnetic ground-state and meta-stable magnetic configurations by identifying 
the energy extrema obtained by iterating self-consistent field equations to convergence
starting from magnetic initial conditions that can be classified as either N\'eel antiferromagnetic (AFM), or ferromagnetic (FM), or
nonmagnetic (NM) states.
The calculation indicates that most 2D MPX$_3$ crystals are semiconductors 
with localized magnetic moments that are ordered antiferromagnetically. 
We find that the use of chalcogen atoms with larger atomic numbers tends to yield smaller  
energy gaps between valence and conduction bands.
In the following we present an analysis of the structural and magnetic properties
of representative 3d transition metal MPX$_3$ trichalcogenides, and
discuss their underlying electronic band structures. 
\begin{figure}[htb!]
\begin{center}
\includegraphics[width=8cm]{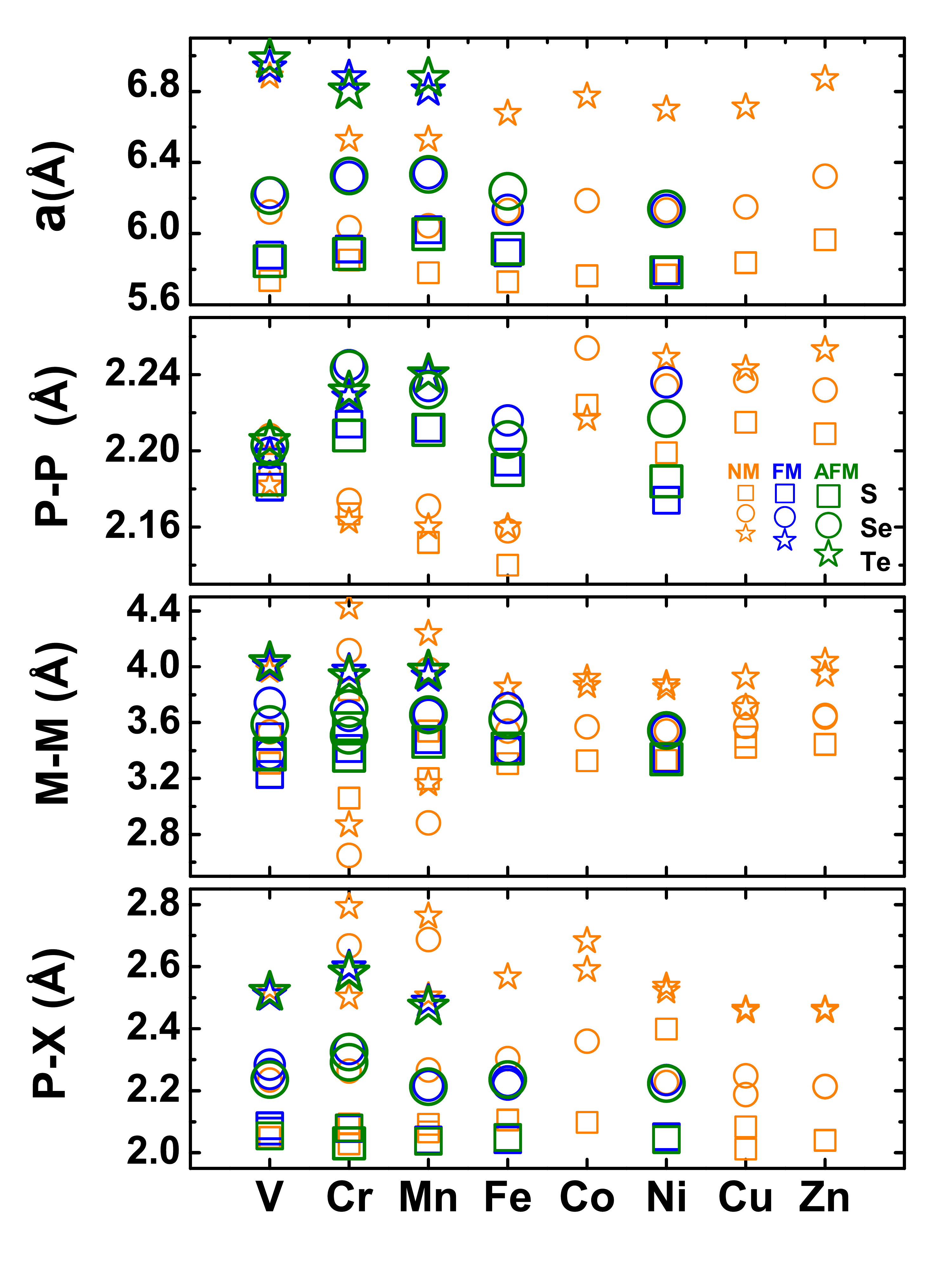}
\caption{(Color online) Summary of key structure metrics
(in-plane lattice constant, P-P bond length, M-M bond length and P-X bond length)
in single-layer MPX$_3$ compounds.
Bond lengths increase with the size of the chalcogen atom (S- square, Se - circle, Te - star)
and a strong systematic dependence on whether the magnetic state is NM (non-magnetic - orange), FM (ferromagnetic - blue), 
or antiferromagnetic (AFM-green), but 
have a more complex non-monotonic dependence on metal atomic 
number, }\label{fig:latticecons}
\end{center}
\end{figure}

\subsection{Structural properties}
\label{sec:structure}
The MPX$_3$ compounds that we study consist of V, Cr, Mn, Fe, Co, Ni, Cu,  and Zn transition 
metal atoms with 3d valence electrons, combined with three different chalcogen atoms S, Se, and Te.
We thus expand our study beyond the most common crystals in this class,
the MnPS$_3$, FePS$_3$, CoPS$_3$, NiPS$_3$, ZnPS$_3$ 
thiophosphates whose bulk structure had been explored in past experiments. 
\cite{mpx3struct1,mpx3struct2,mpx3struct3,mpx3struct4,mpx3struct5,mpx3mag1,mpx3mag2,
mpx3mag3,mpx3mag4,mpx3mag5,mpx3mag6,mpx3mag7,mpx3mag8,mpx3mag9,mpx3mag10,
mpx3mag11,mpx3mag12,mpx3inter2,mpx3inter3,mpx3inter4,mpx3inter5}
We have optimized MPX$_3$ lattice structures using the
unit cell shown in Fig.~\ref{fig:cellstructure} for single layers and assuming ABC stacking for bulk.
Given this framework, structures are characterized by the value of the in-plane lattice constant $a$, the (P-P) 
distance between the phosphorous atoms, the (M-M) distance between
metal atoms, and the (P-X) distance between phosphorous atoms and chalcogen atoms.
The variation in the magnitude of these bond lengths as a function of the metal and chalcogen atom species,
and magnetic configuration is illustrated in Fig.~\ref{fig:latticecons}, 
and summarized numerically in Table~II in the Supplemental Material. 
Results are presented for all metastable magnetic states.  
The structures have a simple dependence on chalcogen atom size, 
with larger in-plane lattice constants and monolayer thicknesses (P-P distances) for larger chalcogens.
However, the dependence of structure on metal atom atomic number is not straightforward.
Experiments in the bulk MPS$_3$ observed a close correlation between the radius of the metal cation 
and the P-P distance of the thiophosphate bipyramid.\cite{mpx3struct1} 
The non-monotonic variation of the metal cation radius with respect to atomic number 
is reflected in the P-P distance trend. 
However, our calculations show that bond lengths correlate more strongly with magnetic 
state than with metal atomic number.

\begin{table}[htb!]
\caption{Comparison between experimental and theoretical 
bulk lattice parameters, and theoretical two dimensional lattice parameters for 
transition metal phosphorous trichalcogenide MPX$_3$ compounds.
These results were obtained using GGA+D2 and are expressed in $\AA$; 
a is the in-plane lattice constant and c$'$ is the layer thickness,  
{\it i..e} it is the distance between the planes
containing the three chalcogen atoms in a single MPX$_3$ layer. 
The values in parenthesis below the bulk c$'$(\AA) values are the 
interlayer distances, {\it i.e.} 
the vertical distances between the top P atom of one layer and the 
bottom P atom of an adjacent layer.
}
\begin{center}
\begin{tabular}{cc|cc|cc|cc}\\ \hline \hline
 MPX$_3$ &X     & {NM }  &            & {FM }  &       &{ AFM }  &   \\ \hline
                &        &a(\AA)  & c$'$(\AA) &a(\AA) & c$'$(\AA)&a(\AA)    &c$'$(\AA) \\ \hline
               &S       & 5.737 & 3.142 & 5.880 & 3.166 & 5.846 & 3.170  \\
      &S(bulk)      & 5.742 & 3.280 & 5.918 & 3.273 & 5.845  & 3.294 \\
      &                 &           &(2.518)&           &(2.585) &           & (2.655) \\
      &S(Expt)     &          &          &          &          &          &   \\
VPX$_3$ &Se    & 6.123 & 3.266 & 6.230 & 3.204 & 6.214 & 3.318 \\ 
               &Te    & 6.885 & 3.328 & 6.935 & 3.418 & 6.980 & 3.431 \\ \hline
               &S      & 5.851 & 3.071 & 5.914 & 3.032 & 5.885 & 3.336 \\ 
      &S(bulk)      & 5.862 & 3.116 & 5.913 & 3.185 & 5.934 & 3.301  \\
      &                 &           &(2.624)&           &(2.506) &           & (2.465) \\
      &S(Expt)     &          &          &          &          &          &    \\
CrPX$_3$  &Se & 6.035 & 3.367 & 6.320 & 3.100 & 6.325 & 3.235 \\ 
               &Te    & 6.530 & 3.672 &  6.880 & 3.330 & 6.804 & 3.388 \\ \hline
                &S    & 5.780 & 2.952 & 6.023 & 3.256 & 5.997 & 3.246  \\
      &S(bulk)     & 5.787 & 2.875 & 6.018  &3.125  & 5.993 & 3.268  \\
      &                 &           &(2.456)&           &(2.567) &           & (2.562) \\
      &S(Expt)    &         &           &           &          & 6.076 &   \\
MnPX$_3$ &Se & 6.045 & 3.182 & 6.340 & 3.416 & 6.334 & 3.402  \\
                &Te   & 6.530 & 3.444 & 6.805 & 3.550 & 6.874 & 3.535  \\ \hline
                &S     & 5.730 & 2.808 & 5.891 & 3.150 &  5.958 & 3.180  \\
      &S(bulk)      & 5.736 & 2.969& 5.874 & 3.313 & 5.921  & 3.217 \\
      &                 &           &(2.386)&           &(2.840) &           & (4.368) \\
      &S(Expt)     &         &          &          &           & 5.934 &   \\
FePX$_3$ &Se  & 6.130 & 2.911 & 6.134 & 3.347 &6.239 & 3.214  \\
                &Te   & 6.676 & 3.038 &          &          &         &   \\ \hline
                 &S    & 5.763 & 2.790 &          &          &         &   \\
      &S(bulk)      & 5.784 & 2.950  &          &          &          &   \\
      &                 &           &(4.250)&           &           &          &  \\
      &S(Expt)     &5.901  &          &          &          &          &   \\
CoPX$_3$ &Se  & 6.186 & 2.850 &          &          &         &    \\
                &Te    & 6.772 & 3.057 &          &          &         &   \\ \hline
               &S      & 5.766 & 2.974 & 5.792 & 3.064 & 5.783 & 3.042  \\
      &S(bulk)      & 5.763 & 2.995 & 5.789 & 3.062 & 5.779 & 3052 \\
      &                 &           &(2.431)&           &(2.449) &           & (2.480) \\
      &S(Expt)     &          &          &          &          & 5.813 &   \\
NiPX$_3$ &Se   & 6.132 & 3.086 & 6.134 & 3.100 & 6.140 & 3.142 \\
                 &Te  & 6.700 & 3.208 &          &          &         &   \\ \hline
                &S     & 5.837 & 3.439 &          &          &         &   \\
      &S(bulk)      & 5.843 & 2.245 &          &          &          &   \\
      &                 &           &(3.903)&           &           &          &  \\
      &S(Expt)     &          &          &          &          &          &   \\
CuPX$_3$ &Se  & 6.151 & 3.532&          &          &         &   \\
                &Te   & 6.711 & 3.617 &          &          &         &   \\ \hline
                 &S    &5.966  & 3.230 &          &          &         &   \\
      &S(bulk)      & 5.973 & 2.564 &          &          &          &   \\
      &                 &           &(3.025)&           &           &          &  \\
      &S(Expt)     &5.971 &           &          &          &          &   \\
ZnPX$_3$ &Se  & 6.323 & 3.379 &          &          &         &   \\
                &Te   & 6.871 & 3.597 &          &          &         &   \\ \hline \hline
\end{tabular}  \label{tab:bulklatticecons}

\end{center}
\end{table}
P-P distances range between 2.15$-$2.3~$\AA$ while M-M bond lengths vary by about 5\% .
The M-M bond lengths within an hexagon 
can be unequal, distorting the hexagonal lattice arrangement of the 
metal atoms as illustrated in Fig.~S1 in the Supplemental Material\cite{SI}.
In general ferromagnetic states have the largest lattice constants, and 
antiferromagnets have intermediate lattice constants.  
The fact that the structural properties of these crystals are correlated with
their magnetic configurations suggests the possibility of controlling magnetic 
properties by straining the lattice, as we will discuss more in detail in a later section.

The comparison between lattice parameters calculated with GGA+D2 and the experimental bulk structure is summarized in Table~\ref{tab:bulklatticecons}. 
The comparison of the GGA and LDA lattice parameters in Table I of the Supplemental Material indicates important discrepancies between the GGA and LDA, 
and is consistent with 
the tendency of the LDA to overbind covalent bonds.  The stronger LDA bonds 
leads to a global reduction of lattice constant and to a stronger distortion of the metal honeycombs.
Within the LDA we find optimized lattice parameters that are  
(2-5$\%$) shorter than in bulk experiments.  The GGA agreement is good (0.5-1.6\%)
for MnPS$_3$, FePS$_3$, CoPS$_3$, NiPS$_3$ and ZnPS$_3$.

\subsection{Magnetic properties}

\begin{figure}[htb!]
\begin{center}
\includegraphics[width=\columnwidth]{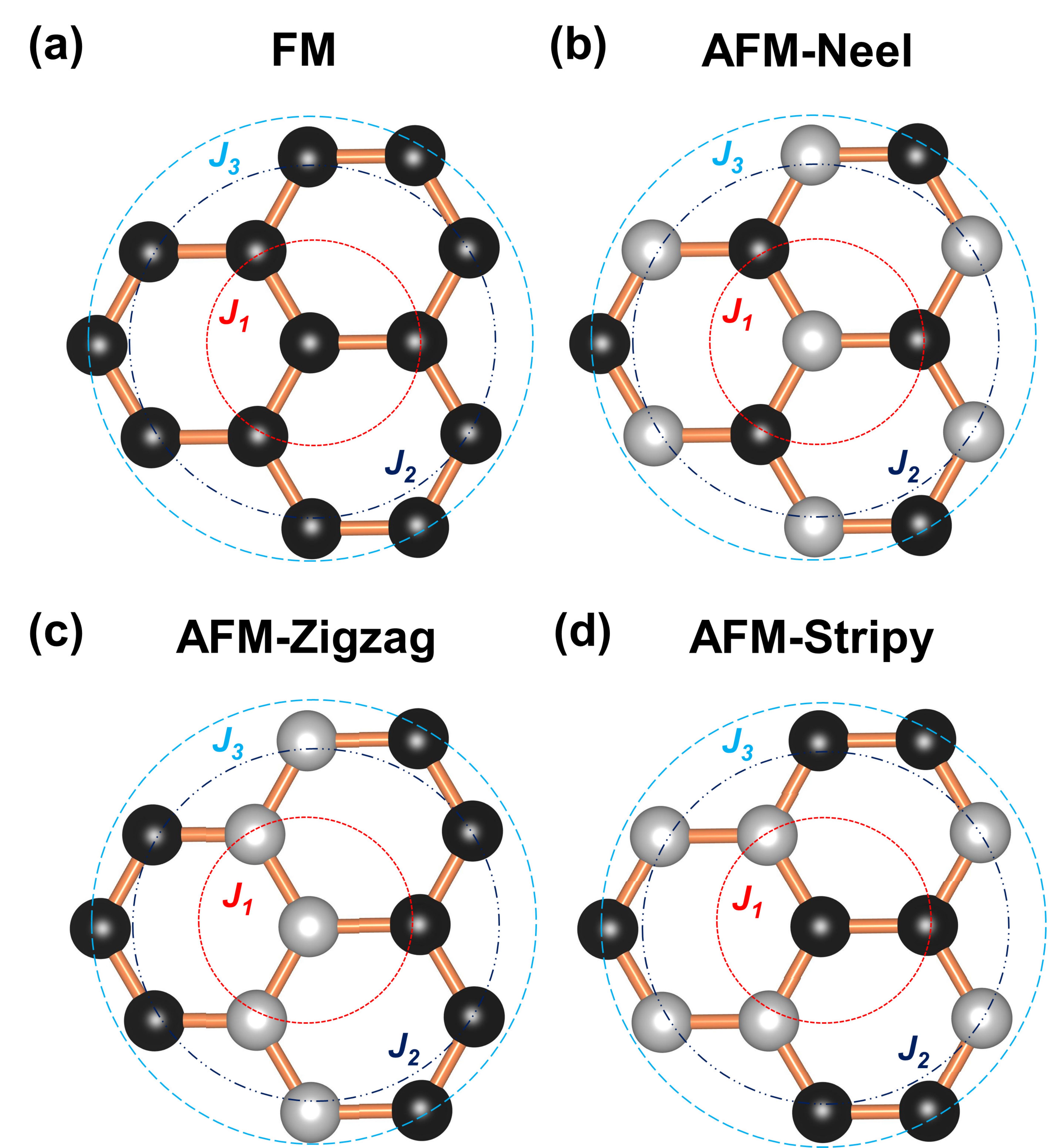}
\caption{(Color online) Top view of different magnetic ordering arrangements
(a) ferromagnetic, (b) N\'eel antiferromagnet, (c) zigzag antiferromagnet and (d) stripy antiferromagnetic.
In MPX$_3$ compounds, the magnetic moments reside primarily on the metal atoms which 
have a honeycomb structure in each layer.  The black and white spheres represent up and down spins, respectively. 
The red circles identify nearest neighbors (NN) of the central metal site, 
the navy blue circles (dashed-dotted ) identify the second nearest neighbors, and the light blue 
circles identify the third nearest neighbors.  The magnetic energy landscape can be approximated by 
assigning Heisenberg coupling constants ${\it J_1}$, ${\it J_2}$ and ${\it J_3}$ to metal atom pairs with
these three separations.} \label{fig:magneticorder}
\end{center}
\end{figure}
\begin{table*}[htb!]
\caption{Total energy relative to that of the lowest energy magnetic configuration
calculated using the GGA+D2 approximation for 
the ferromagnetic (FM), antiferromagnetic (AFM) and non-magnetic (NM) states
of single layer transition metal phosphorous trichalcogenides MPX$_3$.  The absence of an entry for 
a magnetic configuration means that the corresponding state is not metastable.  Energies are expressed 
in terms of meV/unit cell.
Our first principles calculations suggest stable magnetic phases for V, Cr, Mn, Fe, Ni metal compounds.
{The rectangular supercells and structural triangular unit cells used in these 
calculations are illustrated in Fig.~\ref{fig:cellstructure}. The smaller triangular unit cell
is used to compare the NM, FM and Neel AFM phases and the larger rectangular
unit cell is needed when we also compare with the total energies of zAFM, sAFM phases.
The work functions are in eV units for the smaller triangular unit cell.}
}
\begin{center}
\begin{tabular}{cc|ccccc|ccc|ccc}\\ \hline \hline
 MPX$_3$   & X &   \multicolumn{5}{c|}{Rectangular} & \multicolumn{3}{c|}{Triangular}  &  \multicolumn{3}{c}{Work function (eV)} 
  \\ \hline
                   & & NM & FM & AFM & zAFM & sAFM &  {NM }& {FM }&{ AFM }& NM & FM & AFM \\ \hline
                  & S  & 1142.8 & 643.2 & 0.0 &246.9 & 132.7 &  1083.8 & 695.8 & 0.0 &  3.56 & 3.26  & 3.42 \\
VPX$_3$  & Se& 906.9 & 275.8 & 0.0 &164.6 & 87.78 &  1348.1 & 458.8 & 0.0   & 2.82 & 3.57 & 3.58 \\
                  & Te&0.0  &399.3  &323.9  &371.4 & 327.9 &  1296.5 & 183.1 & 0.0    &  3.08 & 3.88 &  3.85 \\ \hline
                  & S  &1454.5 & 0.0 & 91.65 &30.62 & 63.23 &  1182.5 & 0.0 & 73.68  & 3.40 & 3.07 & 3.62 \\
CrPX$_3$  & Se& 1514.6 & 0.0 & 146.1 &57.06 & 522.7 &  1293.7 & 0.0 & 315.4  & 3.98 & 3.52 & 3.65 \\
                  & Te& 1373.4 & 0.0 & 242.9 &81.31 &12.06 & 1482.6 &0.0  & 264.2   & 3.74 & 3.55 & 3.41\\ \hline
                  & S  & 1576.2 & 186.3 & 0.0 &60.68 & 74.59 &  1964.1 & 205.4 & 0.0  & 3.74 & 4.41 & 3.98 \\
MnPX$_3$ & Se& 1356.6 & 156.7 & 0.0 &48.96 & 68.98 &   1417.3 & 166.4 & 0.0 & 3.98 & 4.07 & 3.91\\
                  & Te&0.0  &  &  & &  &  840.0 & 12.44 & 0.0   & 3.68 & 3.66 & 3.77 \\ \hline
                  & S  & 141.5 & 286.8 & 115.8 &0.0 & 289.0 &   44.87 & 0.0 & 308.3   & 3.41 & 4.13 & 4.23 \\
FePX$_3$  & Se& 0.0 &  &  & &  &  0.0   & 165.0 & 324.5   & 3.93 & 3.73 & 3.86  \\
                 & Te & 0.0 &  &  & &  &  0.0   &  &    & 3.32 &  &  \\ \hline
                 & S  & 0.0 &  &  & &  &  0.0    &  &     & 3.84 &  &  \\
CoPX$_3$ & Se& 0.0 &  &  & &  &  0.0   &  &   & 4.33 & &  \\
                 & Te & 0.0 &  &  & &  &  0.0   &  &     & 3.50 &  &  \\ \hline
                 & S   & 507.5 & 276.7 & 64.16 &0.0 & 343.5 &  425.1 &213.4 & 0.0  & 4.63 & 4.91&  5.08\\
NiPX$_3$  & Se & 324.8 & 196.2 & 0.0 &383.2 & 278.2 &  168.0 & 158.8 &0.0   & 4.79 & 4.54 &4.44 \\
                 & Te & 0.0 &  &  & &  &  0.0   &  &    & 2.58& &  \\ \hline
                 & S  & 0.0 &  &  & &  & 0.0   &  &   &4.96 &  &   \\
CuPX$_3$ & Se& 0.0 &  &  & &  &  0.0   &  &   & 5.15 & & \\
                 & Te & 0.0 &  &  & &  &  0.0  &  &    & 2.96& & \\ \hline
                 & S  & 0.0 &  &  & &  & 0.0  &  &    & 3.97 &  & \\ 
ZnPX$_3$ & Se& 0.0 &  &  & &  &  0.0 &  &    &4.47 & & \\ 
                 & Te & 0.0 &  &  & &  &   0.0 &  &   & 1.11 & & \\ \hline \hline
\end{tabular} \label{tab:totalenergies}
\end{center}
\end{table*}

Experimental studies of the magnetic properties of bulk Mn, Fe, Co, Ni thiophosphates 
have found antiferromagnetic ground states with N\'eel temperatures 
ranging between 82K$\sim$155K.~\cite{mpx3struct1,wildes}
Here we find that  magnetism persists in single layer MPX$_3$ compounds.
The magnetic moments develop mainly at the localized metal atom sites, 
except for a noticeable spin polarization that develops on the phosphorous and sulphur atom sites 
in the ferromagnetic configuration.

The late 3d transition elements Cr, Mn, Fe, Co and Ni stand out in the periodic table as elements that tend to order magnetically.   
The bonding arrangements of particular compounds can however enhance or suppress magnetism.
In 2D MPX$_3$ crystals, transition metal atoms are contained within phosphorous trichalcogenide  bipyramidal cages,
and have weak direct hybridization with other transition metal atoms.
The exchange interactions between the metal atoms are therefore mainly mediated
by indirect exchange through the intermediate chalcogen and P atoms.
Magnetic interactions can be extracted from {\it ab initio} electronic structure 
calculations by comparing the ground state energies of different magnetic configurations. 
We compare the energies of
antiferromagnetic, ferromagnetic, and nonmagnetic states 
in V, Cr, Mn, Fe, Ni based compounds in Table~\ref{tab:totalenergies}, where we find that the 
AFM phase is normally favored over the FM phase.

Because our calculations show that the magnetic moments are concentrated at the metal atoms sites
we can characterize the magnetic properties of 2D MPX$_3$ compounds
{ 
by mapping the energy landscape to an effective classical spin Hamiltonian on a honeycomb lattice:
\begin{eqnarray}
H =  \sum_{\langle ij \rangle} J_{ij} \, \vec{S}_i \cdot \vec{S}_j = \frac{1}{2}  \sum_{ i \neq j } J_{ij} \, \vec{S}_i \cdot \vec{S}_j
\end{eqnarray}
where $\vec{S}_i$ is the total spin magnetic moment of the atomic site $i$, 
$J_{ij}$ is the exchange coupling parameters between two local spins,
and the prefactor 1/2 accounts for the double-counting.
The estimated magnetic anisotropy energies that we obtained from non-collinear magnetization 
calculations are on the order of $\sim$160~$\mu$eV per formula unit
for the FM compounds CrPS$_3$, CrPSe$_3$ and on the order of $\sim$1000~$\mu$eV per formula unit for 
the AFM compounds MnPS$_3$, MnPSe$_3$ with magnetization favored perpendicular to the plane.
For this reason, in making the $T_c$ estimates that we describe below, we take the Ising limit of 
this spin-Hamiltonian.
}
By evaluating the three independent energy differences between the four magnetic 
configurations~\cite{mpx3struct1,chaloupka,sivadas} illustrated in Fig.~\ref{fig:magneticorder}, 
ferromagnetic (FM), N\'eel (AFM),  zigzag AFM (zAFM), and stripy AFM (sAFM),
and assuming that the magnetic interactions are short range, we can extract the nearest neighbor
($J_1$), second neighbor ($J_2$), and third neighbor ($J_3$) coupling constants:\cite{sivadas}
\begin{eqnarray}
E_{\rm FM}-E_{\rm AFM} &=& 3 (J_1+J_3) \,  \vec{S}_{A} \cdot \vec{S}_{B}   \\
E_{\rm zAFM}-E_{\rm sAFM} &=& (J_1 - 3J_3) \,   \vec{S}_{A} \cdot \vec{S}_B  \\
E_{\rm FM}+E_{\rm AFM} &-& E_{\rm zAFM}-E_{\rm sAFM} = 8  J_2 \,  \vec{S}_{A} \cdot \vec{S}_A 
\end{eqnarray}
where $\vec{S}_{X}$ is the average spin magnetic moment on the honeycomb sublattice $X$.  
The total energies of the zAFM and sAFM magnetic configurations,
which must be calculated using a rectangular supercell, 
usually have higher energies than the average of the AFM solutions and FM solutions. 
The larger rectangular unit cell required for describing more complex zAFM and sAFM spin configurations 
imposes certain symmetries that restricts the relaxation of the lattices when compared to the triangular unit cell.
 For the case of FePS$_3$, however,  the zAFM has the lowest energy.
The average magnetic moment $S$ at each lattice site 
obtained within the GGA+D2 and GGA+D2+U are listed in Table~\ref{tab:magnetization}.

Single-layer magnetic ordering temperatures T$_{\rm c}$ were estimated by running 
Monte Carlo simulations of the three-coupling-constant effective models
using the Metropolis algorithm in lattice sizes up to $N$=32$\times$64 with periodic boundary conditions, \cite{metropolis,newman,landau}
and verified against calculations performed using Wang-Landau Monte Carlo sampling algorithm in smaller $N$=8$\times$16 lattices.\cite{wanglandau}
We calculated the heat capacity $C= k \beta^2 \left( \langle E^2 \rangle - \langle E \rangle^2 \right)$ as a function of temperature and identified its diverging point as the Neel and Curie temperatures.
See the Supplemental Material\cite{SI} for the plot of the representative results 
for the temperature dependent heat capacity.

\begin{table*}[htb!]
\caption{
Magnetic moments in Bohr magneton $\mu_{\rm B}$ per metal atom for bulk and single 
layer magnetic MPX$_3$ structures, 
the three nearest neighbor exchange coupling strengths $J_i$ in meV implied 
by the Heisenberg model mapping, 
and Monte Carlo estimates of the single-layer critical temperatures based on the 
Ising limit of the classical spin model.  
{The critical temperature is identified with the maximum in the 
derivative of magnetization with respect to temperature.  The Ising limit estimate is motivated by the strong perpendicular 
anisotropy of these materials.} 
The {\it ab initio} calculations were performed 
within GGA+D2, as well as GGA+D2+U using a constant site repulsion at the metal atoms  of U=4eV.
{ 
The magnetization is calculated as the difference between up and down local spin resolved charge densities 
as implemented in Quantum Espresso.
}
}
\begin{center}
\begin{tabular}{cc|cc|cccc|ccc|c|cccc|ccc|c}\\ \hline \hline
 MPX$_3$   & X  & {Bulk} & & \multicolumn{8}{c|}{GGA+D2  }  & 
 \multicolumn{8}{c}{GGA+D2 + U }    \\ \hline
                  &        & {FM}      & {AFM }&{ FM }& {AFM}&{ zAFM }& {sAFM}&{$J_{1}$}&{$J_{2}$}&{$J_{3}$}&{T$_{\rm c}$}  
                  &{ FM }& {AFM}&{ zAFM }& {sAFM}&
                  {$J_{1}$}&{$J_{2}$}&{$J_{3}$}&{T$_{\rm c}$}
                   \\ \hline
                  & S     &  2.03  & 2.00 & 1.92  &2.01&1.68 &1.76& 24.5  &  4.26  &   3.24 & 1100 &  2.96 & 2.93 & 2.95 & 2.93 &  3.93 & 0.147 & 0.143 &    570\\ 
VPX$_3$  & Se   &               &  & 2.02   &2.13&1.77&2.14&  10.2  & 0.340 & 0.437  &  760  & 2.56  & 2.99 & 3.01 & 2.98 &  2.81 & 0.132 & 0.170 & 400 \\ 
                  & Te   &              &  & 2.22  &2.16&2.06&2.19&    9.63 & -3.59  & -7.07  &  570   & 3.10 & 3.08 & 3.13 & 3.08 & 5.10 & -1.68 &-3.51 & 560 \\  \hline 
                  & S     & 2.83  & 3.07 & 2.85  &3.14&2.37&2.34&  -1.73 & -0.015  & 0.029 &  280 & 3.94 & 4.03 & 3.93 & 4.11 & 1.34 & -0.064 & -0.490 & 130 \\ 
CrPX$_3$  & Se   &              &  &2.77   &2.83&2.45&2.38& -9.82 & -3.49  &  6.62 & 1080  &  4.14 & 4.03 & 4.04 & 4.11 & 0.608 & -0.254 & -3.05 & 250  \\ 
                  & Te   &              &  & 2.92  &2.88&2.82&2.48&  -4.89 & 0.006 & -0.087  &  750  & 4.25  & 4.17 & 4.17 & 4.21& 0.849 & 0.333 & -2.10 & 900\\  \hline 
MnPX$_3$                  & S    &  4.24 & 4.15 &  4.26 & 4.16&3.82&3.83&  1.21 & 0.180 &  0.536  &  500  & 4.56  & 4.54 &4.54 & 4.55  &  0.369 & 0.0284 & 0.152 & 200 \\ 
  & Se  &              &  & 4.25  &4.20&4.24&3.82& 0.958 & 0.136 & 0.506  &   450  & 4.56 & 4.54 & 4.55 & 4.56 &0.231 & 0.021 & 0.141 & 150 \\   \hline
 FePX$_3$              & S     &  3.32  & 3.14 & 3.33  & 3.20&3.24&2.48&  -1.59 & -0.179  & 3.00 &  690    &  3.48 & 3.46 & 3.48 & 3.47  & 0.575 & 0.140 & 0.69& 260 \\    \hline
NiPX$_3$                  & S     & 1.25   & 1.13 & 1.26  & 1.14&1.12&1.20&  -11.3  & -0.115  & 36.0 &  1110 & 1.21 & 1.17 & 1.22 & 1.19 & -4.11  & 1.95&17.4  & 560 \\ 
  & Se    &             &  & 0.54  &0.94&0.12&0.99& 127&  -82.3  &  -67.6 &  2580 &  1.14 & 1.08 & 1.22 & 1.14 & -3.11  & 1.04  & 18.2  &  510 \\ \hline
                  \hline
\end{tabular}  \label{tab:magnetization}

\end{center}
\end{table*}

The calculated average values of the magnetic moments in the AFM configuration vary widely, 
assuming the values 
2.01, 3.14, 4.16, 3.2, 0.8 and 1.14 $\mu_B$ for V, Cr, Mn, Fe, Co and Ni based sulphides. (See Table~\ref{tab:magnetization}). 
The magnitudes of the magnetic moments at the metal atoms generally have relatively small differences between different magnetic configurations, between 3\%$\sim$10\%, while larger variations are found in CoPS$_3$ and NiPSe$_3$.
Within GGA+D2 approximation we find that the magnetic moments in 2D MPX$_3$ develop almost entirely at the metal atom sites.
In comparison the phosphorous and chalcogens in the bulk structure do acquire a small spin polarization that can be attributed to interlayer coupling. 
The use of onsite U introduces a small spin polarization enhancement at non-metal atom sites.

\subsection{Band structure and density of states}
Understanding the electronic properties of 2D MPX$_3$ layers is an essential stepping stone on the path toward possible integration in nanodevices.
For spintronic applications of magnetic 2D materials, which seek to couple charge and spin degrees of freedom,
it is desirable to understand how the electronic structure depends on the type of magnetic order.
The spin resolved band structures of the lowest energy spin configuration, calculated within the GGA+D2 approximation, are
presented in Fig.~\ref{fig:bandstructure}, and the associated densities-of-states are presented in Fig.~\ref{fig:dos},
and the projected partial density of states (PDOS) results for various MPX$_3$ compounds presented in Fig.~\ref{fig:pdos}.
Single layer MPX$_3$ with M = V, Cr, Mn, Fe, Co, Ni, Cu, Zn transition metal atoms 
and X = S, Se, Te chalcogen atoms are considered.  
Band structures for higher energy spin configurations are included in the Supplemental Material\cite{SI}. 

\begin{table}[htb!]
\caption{Band gaps of MPX$_3$ compounds predicted to have 
antiferromagnetic semiconductor ground states.  
The band gaps are listed in eV energy units and have values that 
depend substantially on the DFT exchange-correlation approximation employed.
Values are listed for GGA+D2, HSE+D2 hybrid functional and GGA+D2+U approximations. 
The magnitude of U was chosen to maximize the band gap.
As expected the HSE+D2 functional yields the largest band gaps.  
}
\begin{center}
\begin{tabular}{cclclclc}\\ \hline \hline
                  &X   & {GGA+D2} & {HSE+D2}  & {GGA+D2+U}    \\ \hline
                  & S  & 1.23        & 2.50           & 2.32 (U = 3.25)  \\
VPX$_3$    & Se & 0.73       &2.10           & 1.67 (U = 3.25)  \\
                  & Te & 0.41       & 1.57           & 1.00 (U = 3.25)  \\  \hline
                  & S  & 1.36       & 3.05     & 2.34 (U = 5.0)  \\
MnPX$_3$ & Se & 0.98      & 2.24      & 1.70 (U = 5.0)   \\
                  & Te &  0.38     & 1.06      & 0.69 (U = 5.0)   \\  \hline
                  & S  & 0.57       & 2.07     & 2.26 (U = 5.3)  \\
FePX$_3$  & Se & 0.20      & 1.48     & 1.19 (U = 5.3)  \\
                  & Te & 0.13      & 1.35     & 0.59 (U = 5.3)  \\  \hline
                  & S  & 0.84      &2.83       & 1.74 (U = 6.45)   \\
NiPX$_3$  & Se & 0.59      & 2.22      & 1.31 (U = 6.45)  \\
                 & Te & -            &0.27       & -  \\  \hline
                 & S   & 0.01     & 1.76       & 0.08 (U = 1.0)   \\
CuPX$_3$ & Se & -          &0.56       & -\\  \hline
                 & S   & 2.19     & 3.11      & 2.23  (U = 5.4)  \\
ZnPX$_3$ & Se & 1.38    & 2.19       & 1.41 (U = 5.4) \\ 
                 & Te & 0.73    & 1.26       & 0.74 (U = 5.4) \\ \hline \hline
\end{tabular} \label{tab:gaps}
\end{center}
\end{table}

\begin{figure*}[htb!]
\begin{center}
\includegraphics[width=18cm]{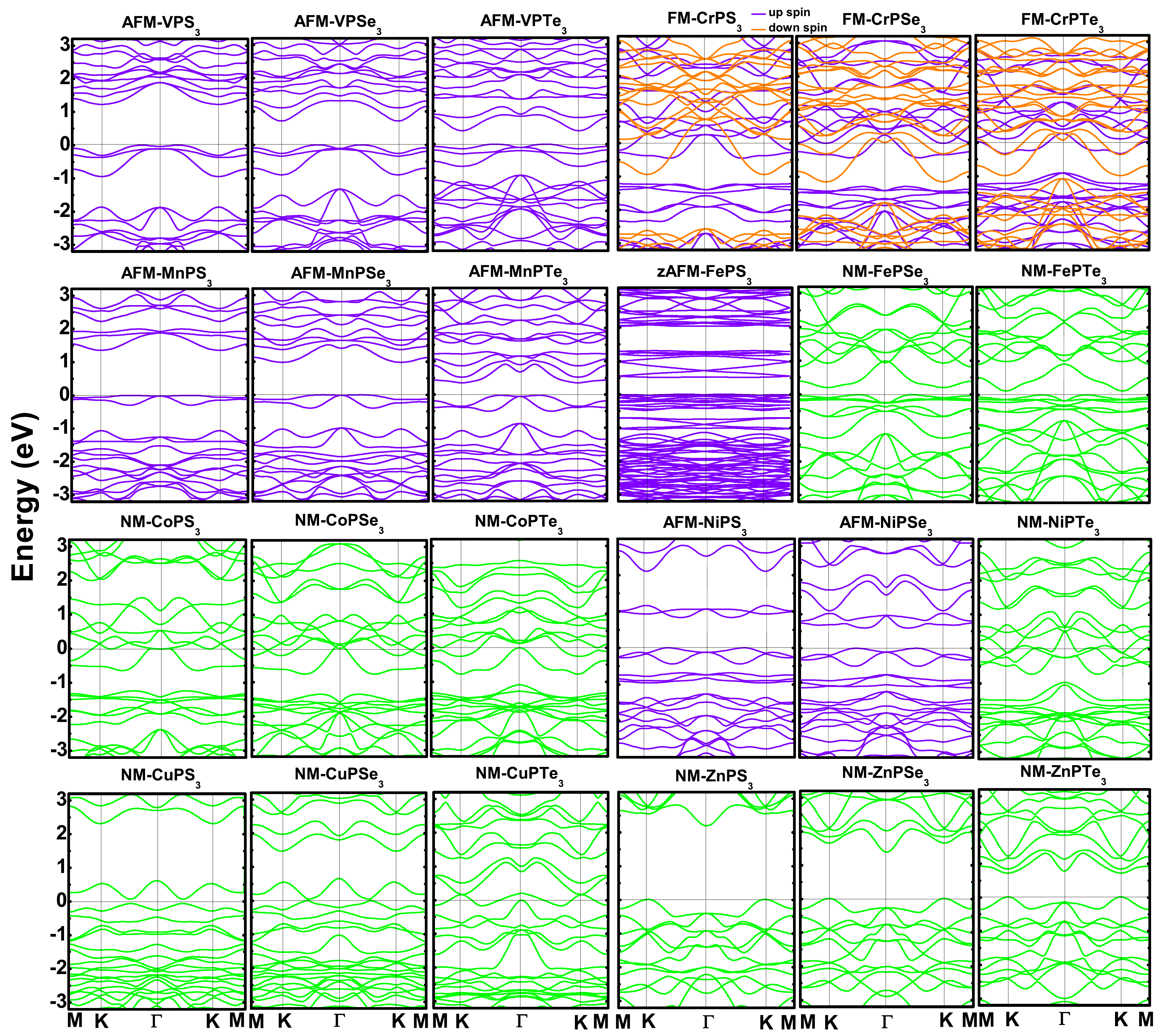}
\caption{(Color online) GGA+D2 band structures for single layer MPX$_3$ compounds 
in their lowest energy magnetic configuration
for M~=~V, Cr, Mn, Fe, Co, Ni, Cu, and Zn  transition metal atoms
and X~=~S, Se, Te chalcogen atoms.  The plotted band structures were calculated using the triangular   
structural unit cell, except
for the case of FePS$_3$ which is predicted to have a larger periodicity zAFM (see text) magnetic structure
that has a triangular unit cell with a doubled lattice constant.
The bands are violet for AFM configurations, violet and orange for the  
up and down split spin bands in FM configurations, and green in NM phases. 
We note that the AFM phases have semiconducting band gaps and that the FM phases are metallic, 
while the non-magnetic phases can be either metallic or semiconducting. 
An overall reduction of the band gaps is observed when we use heavier chalcogen atoms,
in keeping with the reduction of the covalent bond energy in larger shell orbitals. 
} \label{fig:bandstructure}
\end{center}
\end{figure*}
We find that AFMs are almost always gapped semiconductors, 
that the FM states are metallic, and that the NM phases can be either semiconducting or 
metallic.  It follows that the MPX$_3$ class of materials includes almost all
of the behaviors being studied in current spintronics research, including importantly 
both antiferromagnets and ferromagnets and both metals and insulators.  
%
%
%

The AFM band structures for V and Mn based compounds in Fig.~\ref{fig:bandstructure} 
show semiconductor behavior and 
reveal conduction band edges that are near the $K$-points, whereas the valence band edges are between $\Gamma$ and $K$.
This valence bands near the Fermi energy are energetically separated from
deeper lying valence bands. 
We notice from the density of states plot in Fig.~\ref{fig:dos} that there 
is strong sensitivity of the electronic structure to the choice of 
electron-electron interaction model, both for the HSE+D2 approximation containing non-local exchange and 
for on-site repulsion enhancement
introduced through a Hubbard U parameter.  
The analysis of the orbital projected partial density of states in Fig.~\ref{fig:pdos} for the AFM compounds reveals that the conduction 
band edges have an important contribution from the s and p orbitals of the P atom while for the valence band edges 
the chalcogen atom orbitals have an importance presence. 
Other atoms do contribute near the band edges, e.~g. for the V based compounds
the d-orbitals are an important fraction of the valence band edge 
while for the Mn based compounds they lie at deeper energies. 
We can thus expect that surface functionalization and variations in the carrier density at the 
surface chalcogens will have a more immediate impact in the valence bands than the conduction bands. 
The AFM compounds containing Fe and Ni result in non-magnetic solutions when heavier chalcogen atoms are used
within our GGA+D2 calculation. The case of FePS$_3$ has as lowest energy configuration the zigzag AFM phase. 

For MPX$_3$ compounds the FM configurations are metallic as a general rule.
Within our GGA+D2 approximation only the Cr based chalcogenophosphates have FM ground states, 
while they favor the AFM phase when the onsite U is added, see the Supplemental Material for GGA+D2+U results. 
We find that the FM configurations are often meta-stable local minima solutions in MPX$_3$ compounds whose lowest energy solutions are AFM. 
The FM configurations are found in metallic phase at charge neutrality and give rise to half-metallic solutions for MnPS$_3$ and NiPS$_3$.
The half-metallicity could be achieved also in MnPSe$_3$ when carrier doped away from neutrality. 
For the MnPS$_3$ and MnPSe$_3$ compounds we can observe a distinct population of the carriers for the hole and electron sides where doping 
populate orbitals located at the chalcogen and P atoms respectively.

The electronic properties of NM configurations are also presented in Figs.~\ref{fig:bandstructure}-\ref{fig:dos}. 
The NM layers could be potentially interesting if magnetic phases could be induced by forming vertical 
heterojunctions with magnetic materials. 
Unexpectedly, one typically magnetic element (Co) turns out to be non-magnetic in MPX$_3$ compounds 
according to our GGA+D2 based calculations.
As shown in Fig.~\ref{fig:bandstructure} the band structure of these non-magnetic materials can be   
either metallic (Co, Ni, Cu based compounds), semiconducting (Fe), or be a wide gap insulator (Zn).

We now turn our attention to the magnitude of the band gaps of the semiconducting and insulating 2D MPX$_3$ 
chalcogenophosphates calculated within the GGA approximation, 
and using other approximations that partially account for Coulomb correlation through non-local exchange or a Hubbard U, 
are shown in Table~\ref{tab:gaps}.
These gaps are consistently larger when calculated using the 
hybrid HSE+D2 approximation which includes non-local exchange in the energy functional.
A smaller gap enhancement is obtained using a +U interaction correction.  
Band gaps are gradually reduced when S is replaced by heavier chalcogen atoms.

\begin{figure*}[hp!]
\begin{center}
\includegraphics[width=14.5cm]{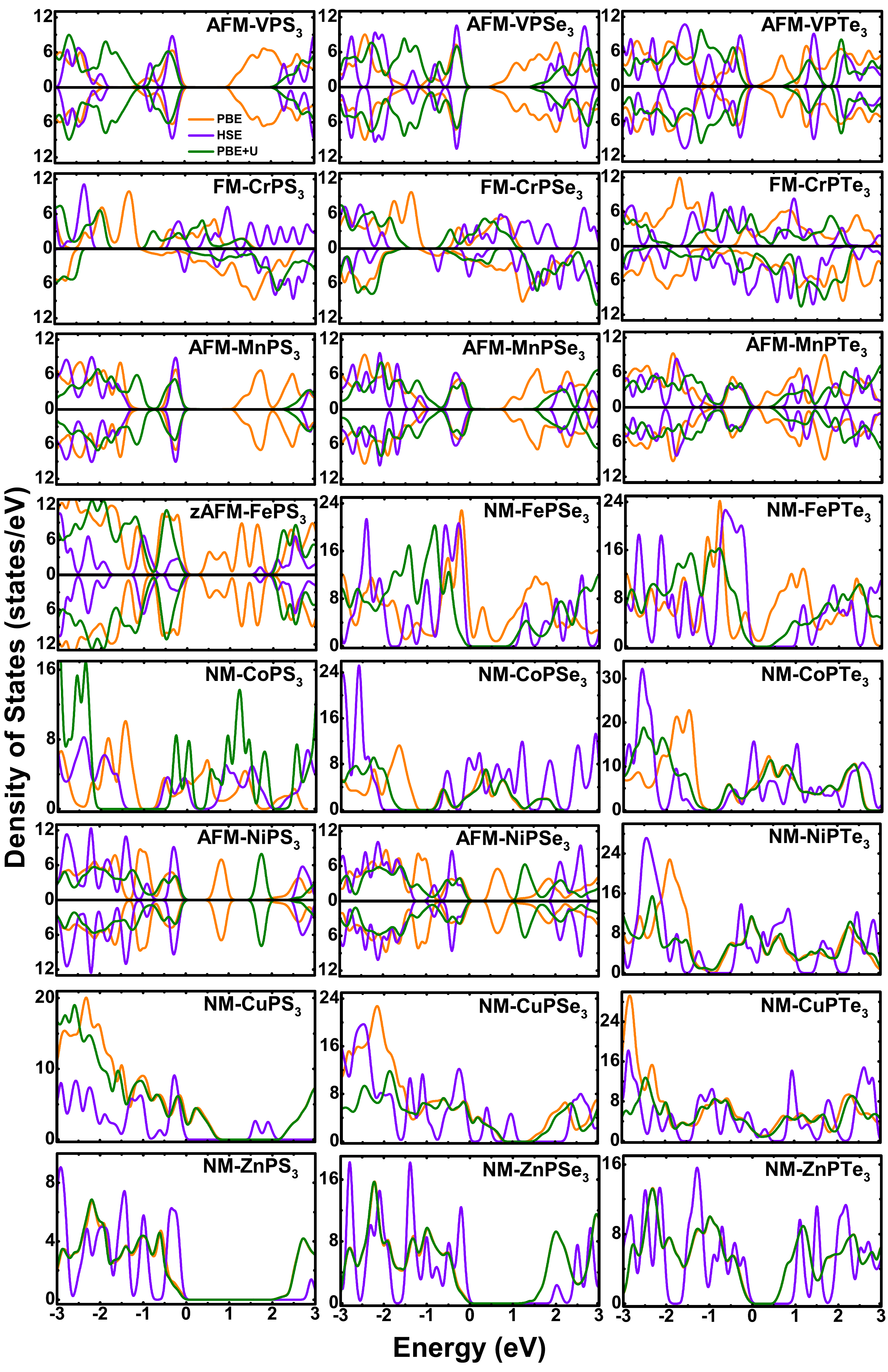}
\caption{(Color online) Total density of states (DOS) for the magnetic ground-states 
in Fig.~\ref{fig:bandstructure}, calculated using three different exchange-correlation energy functional 
calculations { GGA+D2}, HSE+D2 and { GGA+D2}+U.  
We have placed the valence band edge at $E=0$.
Overall band gap enhancements are found for the semiconducting phases when additional electron-electron
interactions are approximated as a non-local exchange corrections in HSE+D2 and as onsite repulsion corrections in 
{ GGA}+U.  Both HSE and { GGA}+U enhance the band gaps.   Other features in the 
DOS plots also depend on the DFT exchange-correlation functional approximation.} 
\label{fig:dos}
\end{center}
\end{figure*}

The corresponding density of states (DOS), plotted in Fig.~\ref{fig:dos}, 
shows the strong influence on the 
ground-state electronic structure when Coulomb correlations are
included.  Note in particular the impact of the non-local Coulomb exchange included in the HSE approximation. 
This suggests that the physics of MPX$_3$ compounds can be dominated by correlation effects
and modeling will be most successful when we rely on effective models that feed from experimental input or
high level {\em ab initio} calculations.

The orbital content of the valence and conduction band edges
that are most relevant for studies of carrier-density dependent magnetic properties
can be extracted from the orbital projected partial density of states (PDOS) results for
various MPX$_3$ compounds presented in Fig.~\ref{fig:pdos}.
Depending on the specific composition and on the magnetic configuration of the material under consideration,
the valence and conduction band edge orbitals can be dominated 
by metal, phosphorous or chalcogen atoms. 
A detailed study on the influence of the Coulomb interaction model on the localization properties of the wave functions
and their influence on the exchange coupling in MPX$_3$ compounds will be presented elsewhere. 

\begin{figure*}[hp!]
\begin{center}
\includegraphics[width=15.5cm]{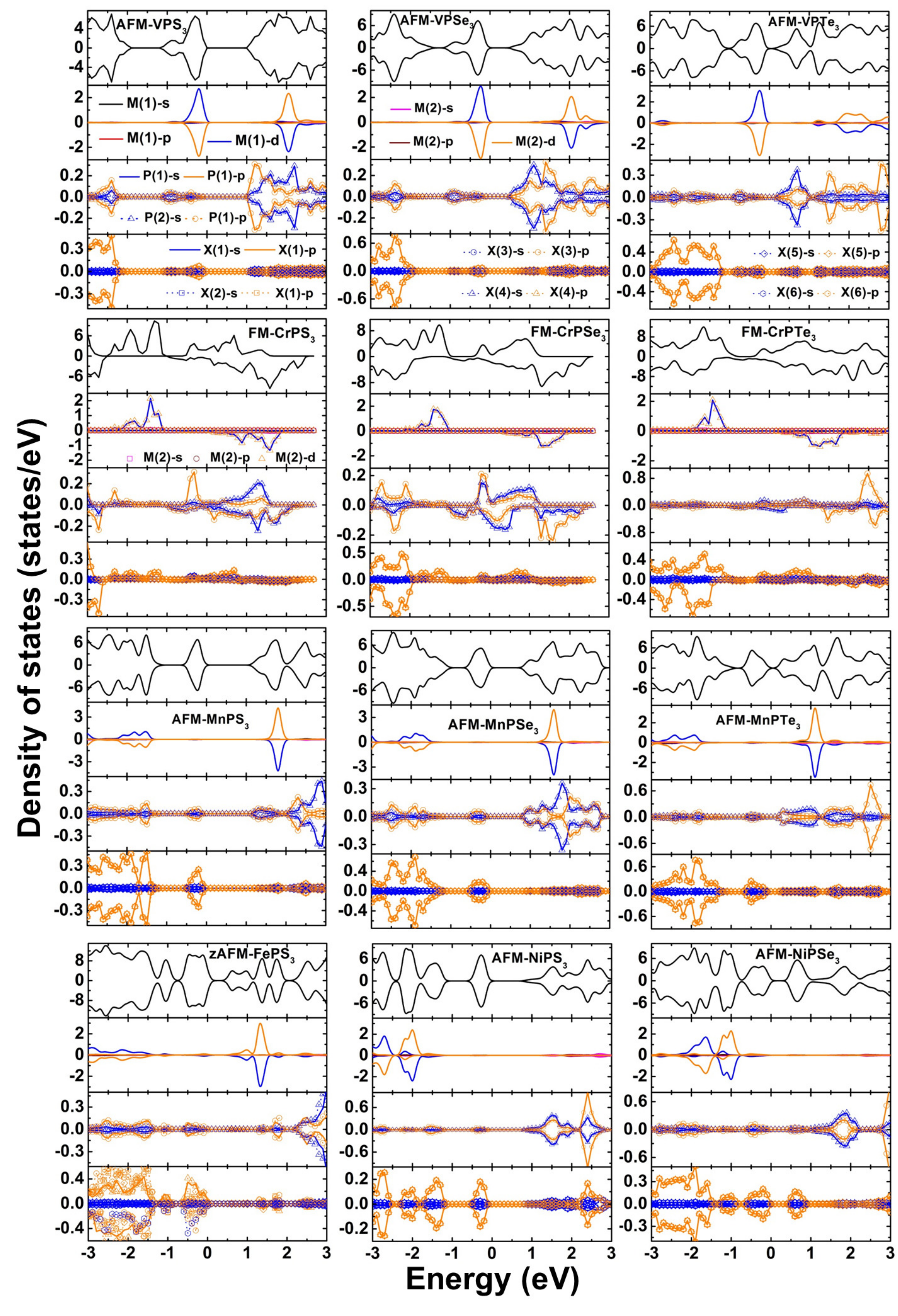}
\caption{(Color online) Orbitally projected partial density of states (PDOS) calculated for self-consistently converged 
ground-state magnetic configurations. The zero of energy is chosen at the valence band edge.
From the PDOS of the various MPX$_3$ compounds we observe that the orbital content of the valence and conduction band edges
vary widely. 
For the compounds showing carrier-density dependent magnetism the states
closest to the Fermi energy generally have an important contribution from the p-orbitals of the 
chalcogen and P atoms. 
 } \label{fig:pdos}
\end{center}
\end{figure*}

\begin{figure*}[htbp!]
\begin{center}
\includegraphics[width=15cm]{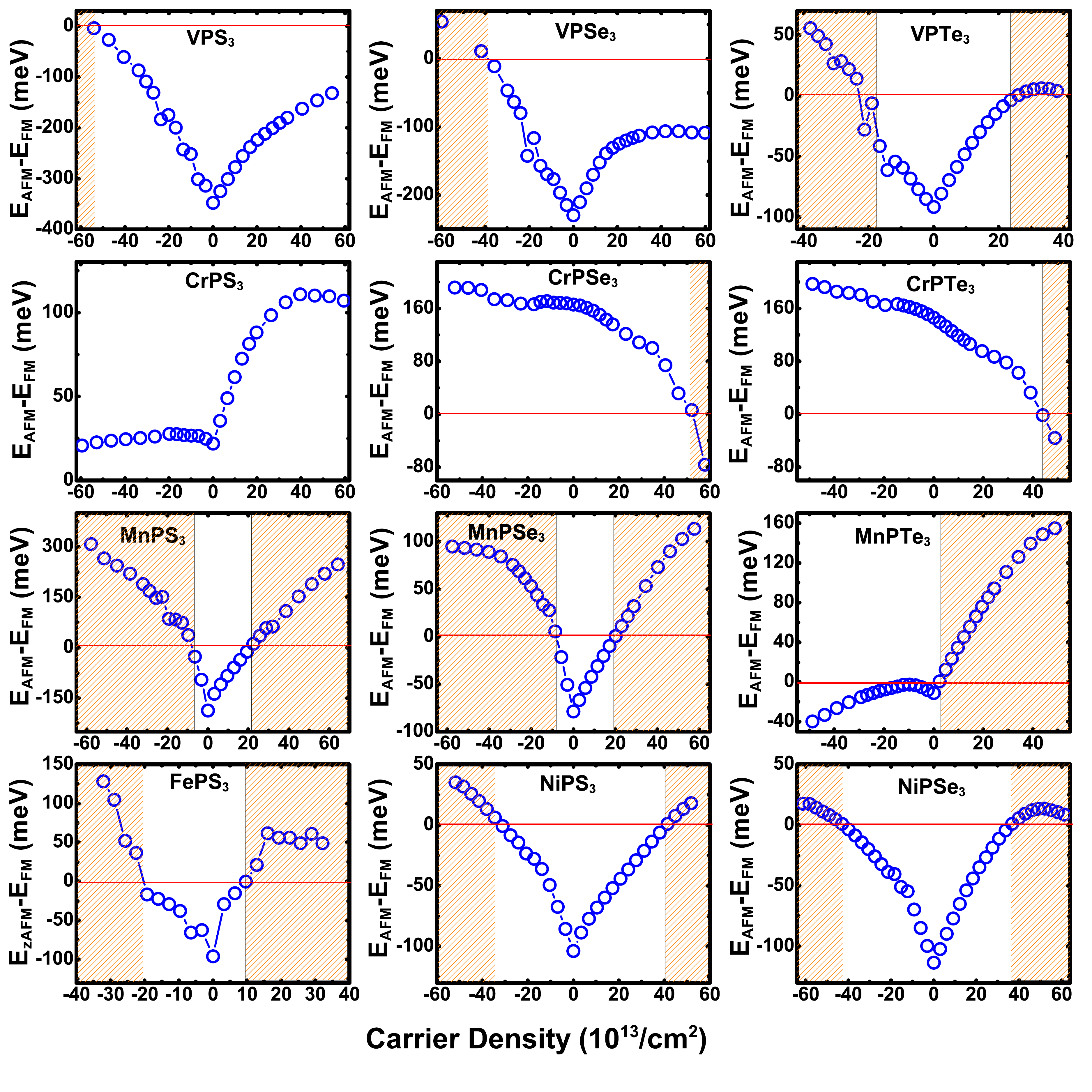}
\caption{(Color online) 
Carrier density dependent total energy differences per MPX$_3$ formula unit 
between the AFM and FM states of V, Cr, Mn, Fe, Ni based single layer trichalcogenides.  
The prominent cusps in these curves appear because 
the antiferromagnetic state has a band gap whereas the ferromagnetic state does not have a band gap.  
The AFM ground-states favored near charge neutrality can be switched to FM states at { accessible}   
carrier densities in V, Mn, Ni, and Fe based compounds.
For CrPX$_3$ with X = Se or Te, it is possible to switch from a FM to an AFM state at 
finite n-doping. 
{ Carrier densities of up to { a few} $ { \sim10^{14}}$ electrons per ${\rm cm}^2$ should be in principle accessible through 
ionic liquid or gel gating.
A carrier density of 0.1 electrons per MPX$_3$ formula unit corresponds to $\sim$$ {6\times 10^{13}}$ electrons per cm$^2$ 
when the lattice constant is $\sim$6$\AA$.}
}
\label{carrier_magnetism}
\end{center}
\end{figure*}

\section{Tunability of magnetic properties}
\label{sec:tunability}
Two dimensional magnetic materials are of interest primarily because of the prospect that 
their properties might be more effectively altered by tuning parameters that are available {\it in situ}. 
Two potentially important control knobs that can be exploited experimentally
in two-dimensional-material based nano-devices are 
carrier-density and strain.  The dependence of magnetic properties on 
carrier density is particularly interesting because it provides a convenient route
for electrical manipulation of magnetic properties.  
In this section we explore the possibility of tailoring the electronic and magnetic properties of MPX$_3$ ultrathin layers
by adjusting carrier density or by subjecting the MPX$_3$ layers to external strains.

\subsection{Field-effect modification of magnetic properties}
The goal of modifying the magnetic properties of a material simply by applying a gate voltage 
is a holy grail in magneto-electronics because it could allow magnetically stored information to be 
written at negligible energy cost.  
Electric field control of magnetic order has been demonstrated in a number of materials.
In ferromagnetic semiconductors the mechanism is carrier density variation in thin films
which leads to a modification of the magnetic exchange interaction and magnetic anisotropy.
In thin films of ferromagnetic metals, gate voltages can vary the Fermi level position at the interface, 
which governs the magnetic anisotropy of the metal. \cite{ohno1,ohno2,endo}

We leave to future work a complete microscopic analysis of the response of a 
MPX$_3$ layer, acting as one electrode of a capacitor, to a gate voltage.
Here we capture the most important response by simply examining the dependence of magnetic 
properties on carrier density.  In doing so we neglect the possible role of 
charge polarization within a MPX$_3$ layer.  
Fig.~\ref{carrier_magnetism} summarizes the theoretically predicted trends in 
competition between AFM and FM states in 
2D MPX$_3$ compounds with M = V, Cr, Mn, Fe, Ni and X = S, Se, Te.
We find that when the magnetic ground state is AFM (M= V, Mn, Ni, Fe), 
transitions to FM states are driven by sufficiently large electron or hole carrier densities. 
The origin of this trend is easy to understand.  Because the FM state is gapless
the energy changes associated with adding one electron and removing one electron are  
identical.  In the gapped AFM states they differ by the energy gap $E_{gap}$.  It follows that the 
energy difference per area unit between antiferromagnetic and ferromagnetic states $\delta E \equiv E_{AFM} -E_{F}$ 
{  is given at low carrier densities by 
\begin{eqnarray}
  \delta E(n) &=& \delta E_0  + {  ( E_{gap}/2 +  \delta \mu ) \, n  \nonumber } \\
  \delta E(p) &=& \delta E_0   + {  (E_{gap}/2 - \delta \mu) \, p   }
\end{eqnarray} 
where $n$ is the carrier density of $n$-type samples, $p$ is the carrier density of $p$-type samples, }
$\delta E_0$ is the energy difference per area unit between antiferromagnetic and ferromagnetic states in neutral 
MPX$_3$ sheets, and $\delta \mu$ is the difference between the mid-gap energy of the 
antiferromagnetic semiconductors and the chemical potential of the ferromagnetic metals.  
Introducing $n$-carriers is most effective in driving a transition from antiferromagnetic to 
ferromagnetic states when $\delta \mu$ is positive, whereas introducing $p$-carriers is 
most effective when $\delta \mu$ is negative.  By comparing with Fig.~\ref{carrier_magnetism} we 
conclude that $\delta \mu$ is small in most cases, but large and negative in 
MnPS$_3$ and MnPSe$_3$.
Because the energy difference per formula unit between ferro and antiferromagnetic states is 
much smaller than the energy gap, a transition between antiferromagnetic and ferromagnetic states 
can be driven by carrier density changes per formula unit that are much smaller than one - especially so 
when $\delta \mu$ plays a favorable role.  In particular we see in Fig.~\ref{carrier_magnetism} that a 
transition between ferromagnetic and antiferromagnetic states are predicted in MnPS$_3$ and MnPSe$_3$
at hole carrier densities that are $\sim$$ {10^{14}} {\rm cm}^{-2}$, which corresponds to about $\sim$0.16 electrons per
formula unit.  Density changes of this magnitude can be 
achieved by ionic liquid or gel gating.  Since this size of carrier density is sufficient to completely change 
the character of the magnetic order, we can expect substantial changes in magnetic energy landscapes 
at much smaller carrier densities.   
Transitions between ferromagnetic and antiferromagnetic states are predicted
in most MPX$_3$ compounds.  
Our calculations show that the FM solution is the preferred stable magnetic configuration 
in almost all cases when the system is subject to large electron or hole densities in the 
range above { a few times } ${\pm 10^{14}}$ cm$^{-2}$.  
The DOS and PDOS shown in Figs.~\ref{fig:dos}-\ref{fig:pdos} in the main text and in Figs.~5-10 in the Supplemental Material may suggest that a Stoner ferromagnetic instability is at play when the peaks near the band edges are sufficiently prominent.
Our calculations therefore motivate efforts to find materials which can be used to establish 
good electrical contacts to MPX$_3$ compounds, and in particular to 
the valence bands of MnPS$_3$ and MnPSe$_3$.
For the Cr based compounds, whose behavior is different,
the ground states at zero carrier density is FM and we find 
that a transition to an AFM state can occur for $n$-doped systems.

\begin{figure*}[htbp!]
\begin{center}
\includegraphics[width=14.5cm]{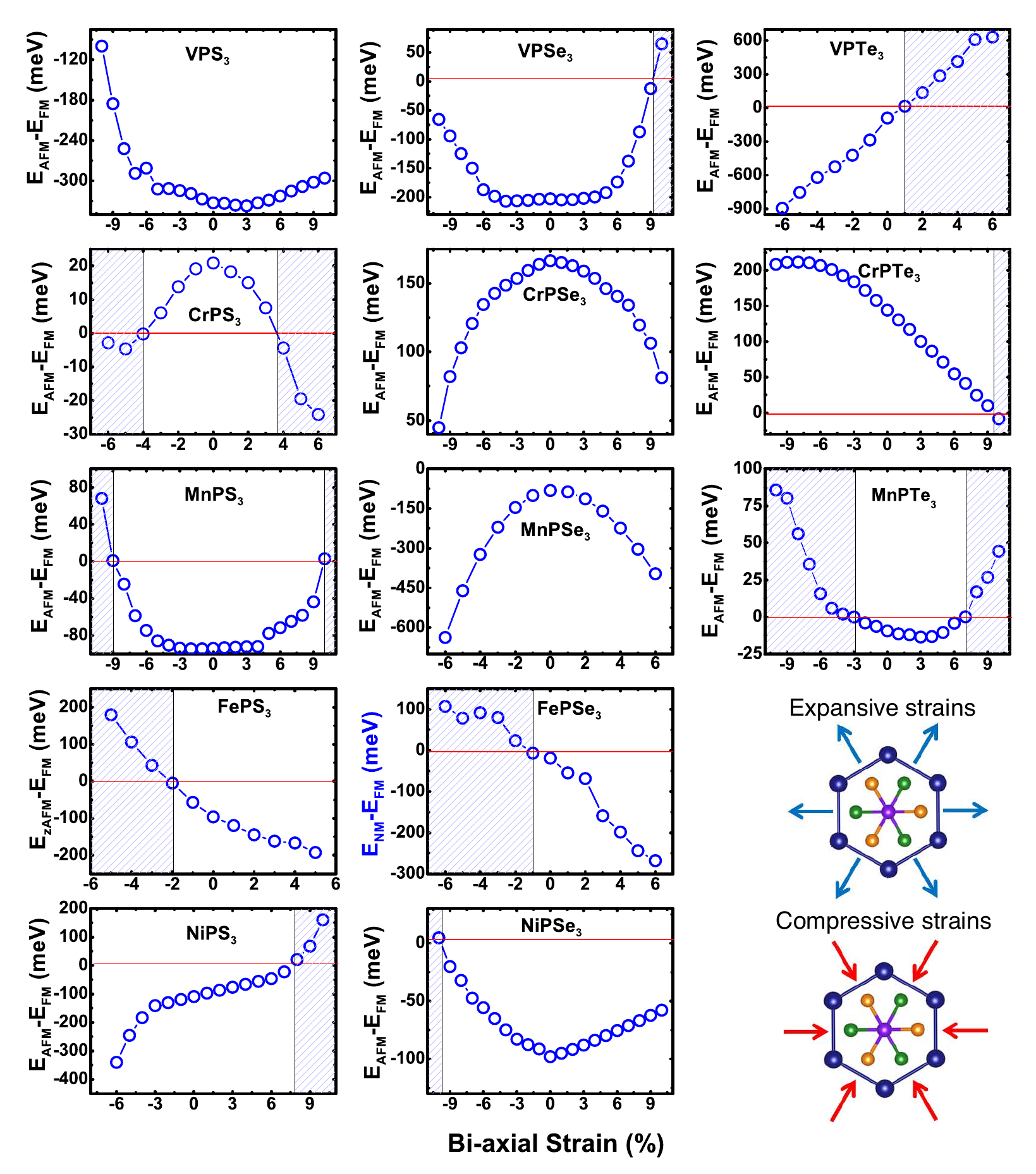}
\caption{(Color online) 
Influence of in-plane strain on the magnetic configurations of selected MPX$_3$ compounds. 
Magnetic phase transitions are introduced by in-plane biaxial compressive and expansive strains for several magnetic 
MPX$_3$ compounds at zero carrier density.  
Particularly sensitive strain dependence is seen for FePS$_3$, FePSe$_3$ and VPTe$_3$
where small strains on the order of 1\% of the lattice constant can switch the magnetic phases. }
\label{fig:strain}
\end{center}
\end{figure*}

\subsection{Strain-tunable magnetic properties}
The membrane-like flexibility of ultrathin 2D materials makes them suitable platforms for 
tailoring the electronic structure properties by means of strains. 
Representative examples on effects of strains in 2D materials properties that have been discussed 
in recent literature include the observation of Landau-level like density of states near high curvature graphene bubbles \cite{guineabubbles},
or the commensuration moire strains that opens up a band gap at the primary Dirac point 
in graphene on hexagonal boron nitride for sufficiently long moire patterns \cite{jarillogap,woods,origingap}.
The sensitive variation of the different M-M, P-P, M-X bond lengths as a function of magnetic configuration
we have presented earlier in Sec.~\ref{sec:groundstate} suggests that strains can be used as switches
to trigger magnetic phase transitions or alter the stability of the magnetic phases by modifying
the total energy difference with respect to the non-magnetic phases.
Here we have carried out calculations of the total energies for different magnetic phases to explore the influence of 
expansive or compressive in-plane biaxial and uniaxial strains and the effects of external pressure along the out-of 
plane axis in 2D MPX$_3$ materials.
The compressive and expansive biaxial strains have been modeled through uniform scaling of the rectangular unit cell 
in Fig.~\ref{fig:cellstructure} and likewise the uniaxial strains are modeled by scaling the unit cell either in the zigzag 
or armchair directions. 
We notice that generally the effects of uniform biaxial versus uniaxial strains introduce modifications in the magnetic phase 
energy difference total energies that are qualitatively similar for the compressive or expansive strains,
and that the effects are stronger for { biaxial} strains when compared to uniaxial strains
which do not show a noticeable difference between the zigzag or armchair directions.
Pressure strains along the out of plane vertical c-axis have been applied by artificially modifying the 
P-P distance and we find for MnPS$_3$ a clear transition from AFM to FM with the total energy 
difference $E_{\rm AFM} - E_{\rm FM}$  { changing from 
$-$93~meV to 268~meV per formula unit}
in the presence of a $-$4\% bond length shortening strain for the P-P distance.
Relatively large strains are required to alter the magnetic properties in compounds like MnPS$_3$ but 
other systems such as 
CrPS$_3$, FeP$S_3$, FePSe$_3$ NiPS$_3$ and VPTe$_3$ the magnetic phase transitions can be achieved for smaller strains
as they have relatively smaller total energy differences between the magnetic phases.
We find a particularly sensitive transition near charge neutrality for FePS$_3$, FePSe$_3$
and VPTe$_3$ where strains on the order of $\sim$1\% is enough to switch between different
magnetic configurations.

\section{Summary and discussion}
\label{sec:summary}
In this paper we have carried out an {\em ab initio} study of the MPX$_3$ transition metal phosphorous trichalcogenide class
of two-dimensional materials.  We have considered different 
combinations of 3d metals (M = V, Cr, Mn, Fe, Co, Ni, Cu, Zn) and chalcogens (X = S, Se, Te)
in an effort to identify materials that are promising for spintronics based on two-dimensional materials.
Our calculations suggest that magnetic phases are common in the single-layer limit of these van der Waals 
materials, and that the configuration of the magnetic state depends systematically on the
transition metal/ chalcogen element combination. 
We find that semiconducting N\'eel antiferromagnetic states  
are most common; and predict AFM phases for V, Mn, Fe, and Ni based compounds.
A metallic ferromagnetic states is found in Cr based compounds, 
and non-magnetic states are found with both semiconducting and metallic electronic structure, 
while introduction of U tends to stabilize the AFM magnetic phase.
As expected on the basis of the weaker covalent bonds of larger atomic orbitals
we find that for semiconducting antiferromagnetic materials, replacing a smaller 
chalcogen atom with larger chalcogen atom
reduces the energy band gap and as a consequence also the difference between
the ground state energies of FM and AFM states.
Interestingly we do not find magnetic states for CoPX$_3$, even though
Co is typically a magnetic atom.

The electronic structures predicted by density functional theory for  
these materials are sensitive to the choice of exchange-correlation energy functional.
For example, there are substantial discrepancies in predicted densities-of-states between 
the standard semi-local GGA, approximations with 
a local U correction, and hybrid functionals with non-local exchange.
This sensitivity of the optimized ground-state results to the choice of the approximation scheme makes it desirable 
to benchmark the results against experiment in order to establish which approximation 
is most reliably predictive.
The fact that the predictions of density-functional theory are qualitatively sensitive to the exchange-correlation
approximation employed, indicates that the MPX$_3$ compounds are 
correlated.

An interesting interdependence between magnetic order and atomic structure  
was found, with typical variations on the order of a few percents in the lattice constant and interatomic bond lengths 
leading to structural distortions depending on the magnetic configuration of the system.
The bond length variations were traced through M-M, P-P, P-X bond distances associated with
the distortions in the honeycomb array of the transition metal atoms 
and the distance between the atomic centers in the (P$_2$X$_6$)$^{4-}$ bipyramids.
We leave for future work the analysis on the role of spin-orbit coupling effects and optical properties 
arising from lattice symmetry breaking coupled to the onset of magnetism.  %
Sizeable differences between the GGA and the LDA solutions is suggestive of the delicate balance between the 
different chemical bonds for configuring the optimized structure of these compounds
which in turn are affected by the onset of magnetism.  In the antiferromagnetic state, the strength of the 
exchange interactions is expected to vary inversely with the band gap; approximations that underestimate 
the band gap will overestimate exchange interaction strengths.  We expect that the LDA likely does 
underestimate band gaps, as it commonly does, and therefore the exchange interactions it implies may
well be overestimates.  
In 2D MPX$_3$ a variety of different stable and meta-stable magnetic configurations were found. 
Semiconducting states with a band gap are typically antiferromagnetic phases in either N\'eel, zigzag, 
or stripy configuration and the metallic states are typically ferromagnetic phases. 
The critical temperatures for magnetic phases in the single layer limit are generally expected
to be smaller than in the bulk due to the reduction in the number of close neighbor exchange interactions,
although changes in the degree of itineracy and shifts in relative band positions can also play a role. 
We analyzed the magnetic phases of the 2D MPX$_3$ compounds by building an effective model Hamiltonian 
with exchange coupling parameters extracted by mapping the total energies from our {\em ab initio} calculations 
onto an effective classical spin model.  The Curie and N\'eel temperatures T$_{\rm C}$, T$_{\rm N}$ 
are obtained by using a statistical analysis based on the Metropolis algorithm \cite{newman}.
The calculated N\'eel temperatures of the antiferromagnetic compounds
have a wide range of variation, ranging between a few to a few hundred Kelvin.

Control of magnetic phases by varying the electric field in a field effect transistor device
is a particularly appealing strategy for 2D magnetic materials.
Our calculations indicate that a transition between antiferromagnetic and ferromagnetic phases
can be achieved by inducing carrier densities in 2D MPX$_3$ compounds
that are in the range that can be induced by a field effect. 
We find that a magnetic phase transition from a FM state to an AFM state can be induced in 
metallic CrPSe$_3$ and CrPTe$_3$ compounds, which are close to the phase boundary, by 
gating to large n-type carrier densities.  Similarly AFM to FM transitions can be achieved in
VPTe$_3$, MnPTe$_3$, and FePS$_3$.  
For materials exhibiting magnetic phases we find that using the heavy chalcogen Te 
can reduce the carrier densities required for magnetic transitions.

The interdependence between atomic and electronic structure
suggest that strains can be employed to tune magnetic phases.  
We have found that the ground state magnetic configuration can undergo phase transitions 
driven by in-plane compression or expansion of the lattice constants or by as c-axis pressure.
The magnitude of the required strains vary greatly, with values below 1\% for 
compounds such as FePS$_3$, FePSe$_3$ and VPTe$_3$,
and larger strains on the order of or far greater than 4\% are required to trigger a FM-AFM 
magnetic phase transition in CrPS$_3$ and NiPS$_3$.
Phase transitions for even larger strains on the order of $\sim$9\% are observed in 
VPSe$_3$, MnPS$_3$, NiPS$_3$, NiPSe$_3$
whose non-strained configuration has a robust gapped antiferromagnetic phase.

Based on our calculations, we conclude that single layer MPX$_3$ transition metal thiophosphates
are interesting candidate materials for 2D spintronics.
Their properties, including their magnetic transition temperatures, can be adjusted by the application
of external strains or by modifying the carrier densities in field effect transistor devices.
The sensitivity of these systems to variations in system parameters, such as composition stoichiometry,
details of the interface, and the exchange coupling of the magnetic properties with external fields,
offers ample room for future research that seek new functionalities.

\section{Acknowledgements}
We are thankful to the assistance from and computational resources provided by the
Texas Advanced Computing Centre (TACC). 
JJ was supported by the 2015 Research Fund of the University of Seoul. 
This work was also supported by NRF-2014R1A2A2A01006776 for EHH, DOE BES Award SC0012670, and Welch Foundation 
grant TBF1473 for AHM, and NRF-2015R1D1A1A01060381 for MH.
We acknowledge helpful discussions with J. D. Noh on the calculation of the critical temperatures of the 
magnetic phases.

\newpage

\begin{center}
{\bf \large Supplementary Information}
\end{center}
We present the lattice structures, band structures, the associated density of states and 
the orbital projected density of states (PDOS) calculated for self-consistently 
converged magnetic configurations with total energy local minima with  
the energy origin E=0 is placed at the valence band edge for gapped bands.
Carrier density dependent magnetic phase transition is calculated using a finite 
onsite repulsion U=4~eV on the of GGA+D2.
We also obtain the magnetization as a function of temperature through the
Metropolis Monte Carlo simulation in a 32$\times$64 superlattice.

\begin{figure*}[htb!]
\begin{center}
\includegraphics[width=17cm]{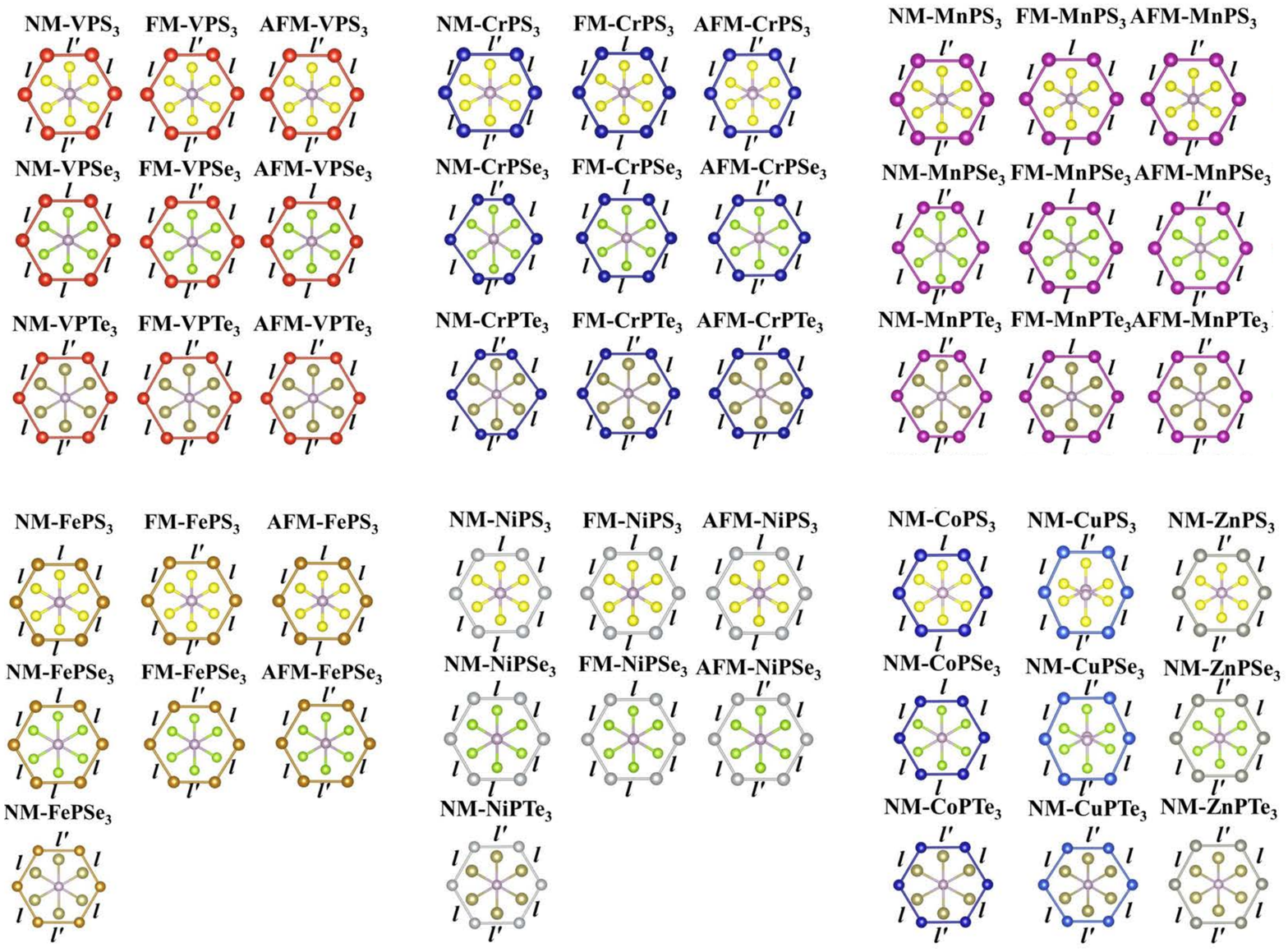}
\caption{(Color online) Illustration of the lattice structures corresponding to the different bond length parameters from Table~I of the main text
for different meta-stable magnetic configurations. 
Unequal metal-metal bond lengths are represented through $l$ and $l'$ which distorts the hexagonal lattice arrangement of the metal atoms.  
Different colors represent different transition metal and chalcogen atoms.
} \label{fig:bandstructure}
\end{center}
\end{figure*}

\begin{table}[htb!]
\caption{Theoretical  lattice parameters in $\AA$ for 
transition metal phosphorous trichalcogenide MPX$_3$ within GGA and LDA;
a is the in-plane lattice constant and c$'$ is a layer thickness parameter, 
{\it i..e} it is the distance between the planes
containing the three chalcogen atoms in a single MPX$_3$ layer. 
}
\begin{center}
\begin{tabular}{cc|cc|cc|cc|cc|cc|cc}\\ \hline \hline
MPX$_3$    &        &   \multicolumn{6}{c|}{LDA}                        & \multicolumn{6}{c}{GGA}  \\ \hline \hline
                  &        & {FM }   &                 & {AFM }  &                 &{ NM }  &                  & {FM }  &                 & {AFM }  &                 &{ NM }  &   \\ \hline
                  & X     &a(\AA)   & c$'$(\AA) &a(\AA) & c$'$(\AA) &a(\AA)    &c$'$(\AA)  &a(\AA)  & c$'$(\AA) &a(\AA) & c$'$(\AA) &a(\AA)    &c$'$(\AA) \\ \hline
                  &S      &5.616    & 3.106    & 5.675    &  3.109    & 5.616   & 3.106          &5.913    &  3.061   &  5.886    & 3.161 & 5.7647  &3.134     \\
VPX$_3$    &Se    & 5.951   &3.230     &6.053     &3.245      &6.001     &3.221          &6.249    &3.234     &6.259      &3.313  &6.167     &3.251        \\ 
                  &Te    &6.452    &3.595     &6.608     &3.418      &6.459     &3.449          &6.409    &3.680     &7.087       &3.417  &6.963     &3.294        \\ \hline
                  &S      &5.716   & 2.922    & 5.771   & 3.004     & 5.745   & 3.028           &5.957    & 3.016     &5.928     & 3.337  & 5.886    & 3.061  \\ 
CrPX$_3$  &Se    &6.186   & 3.018    & 6.170    & 3.111     & 6.181   &3.147           & 6.368    & 3.077     & 6.322  & 3.140    & 6.354    & 3.179\\ 
                  &Te    & 6.654   & 3.285    &6.646    & 3.291    &  6.664    & 2.926         & 6.862    & 3.358   &  6.881   &  3.359  &  6.551   & 3.046\\ \hline
                  &S      &5.667   & 2.865   & 5.869    &  3.138   & 5.691     &  2.877          & 6.086   &3.246      & 6.058   & 3.235   & 5.814    & 2.932 \\
MnPX$_3$ &Se    &6.0916 &2.925    &  6.198   & 3.271    & 6.125     &  2.964          & 6.409   &3.417      & 6.395  & 4.017   & 6.090     & 2.966  \\
                  &Te    & 6.602  &3.123    & 6.712    & 3.041    & 6.625     &  3.014          & 6.841   &3.565      & 6.951  &3.533    &6.554      &3.215   \\ \hline
                  &S      &5.639   &2.702    &5.639   &2.702     & 5.638       & 2.702           &5.914    &3.072     & 5.963     & 3.105  & 5.758    & 2.786 \\
FePX$_3$  &Se    &6.035   &2.800    &6.036   &2.800     &6.035        &2.800            &6.290    &3.294     &6.326      &3.205   &6.180     &2.876   \\
                  &Te    &              &              &               &          &6.545   &2.957                              &              &              &               && 6.740     &2.998        \\ \hline
                  &S      &              &               &              &&5.677   &2.682                           &              &              &              & & 5.801    &2.7571               \\
CoPX$_3$ &Se        &              &              &               &&  6.093  &2.764                                    &              &              &               &&6.241    &2.175         \\
                 &Te         &              &              &               & & 6.572   &2.986                                 &              &              &               & &6.763     &3.027        \\ \hline
                 &S       &5.662  & 2.880    &5.655     &2.876     & 5.655     &2.876          & 5.824   &  3.059   &  5.8147  &  3.036  &5.795 & 2.963        \\
NiPX$_3$  &Se     &6.020  &2.978     &6.020     &2.988     &6.020      &2.978          &6.181    &3.073     &6.181      &3.124    &6.177  &3.060         \\
                 &Te         &              &              &               &&6.553   &3.099                         &             &              &               &  &6.898    &3.030        \\ \hline
                 &S           &              &               &               &&5.711  & 3.313                         &              &              &               & &5.900   &3.474        \\
CuPX$_3$ &Se         &              &              &                && 6.043  &3.278                                 &              &              &               &&6.222   &3.186         \\
                 &Te                  &              &               &                && 6.496  &3.511               &              &              &               & &6.770   &3.347           \\ \hline
                 &S           &              &               &                &  &5.826 & 3.174                       &              &              &               & & 6.022  &3.224          \\
ZnPX$_3$ &Se          &              &                &                &  &6.170 &3.314                               &              &              &               &&6.382  &3.385            \\
                 &Te               &              &                &                &    &6.672  &3.533                        &              &              &               & & 6.898  &3.599       \\ \hline \hline
\end{tabular}  \label{tab:bulklatticecons}

\end{center}
\end{table}

\begin{table*}[htb!]
\caption{Self-consistent theoretical M-M, P-P, and P-X bond lengths in $\AA$  for 
metastable magnetic states of single layer transition metal phosphorous trichalcogenide MPX$_3$ within GGA+D2.
Double entries indicate bond length variations that distort the reference lattice structure.}
\begin{center}
\scalebox{1}{%
\begin{tabular}{cc|ccc|ccc|ccc}\\ \hline \hline
 MPX$_3$ & X   &            & {M-M}  &           &   & {P-P}  &   &      & {P-X}   &  \\ \hline 
                 & & {NM }& {FM }&{ AFM }&{NM }& {FM } &{ AFM } &{NM } &{FM } &{AFM } \\ \hline 
                 &S & 3.313, 3.311 & 3.500, 3.230 & 3.377, 3.372       & 2.193 & 2.181 &2.185      & 2.050 & 2.070, 2.087 & 2.055        \\ 
VPX$_3$  &Se & 3.535 & 3.743, 3.375 & 3.587                           & 2.208 & 2.199 & 2.203     & 2.235 & 2.284, 2.254 & 2.236         \\ 
              &Te & 3.976, 3.974 & 4.002, 4.005& 4.030, 4.028      & 2.182 & 2.198 & 2.205     & 2.526 & 2.507 & 2.518                         \\ \hline 
              &S & 3.060, 3.836 & 3.414 & 3.363, 3.558                 & 2.167   & 2.214 &2.208     & 2.093, 2.029 & 2.078 & 2.080, 2.031     \\ 
CrPX$_3$  &Se & 4.115, 2.650 & 3.648 & 3.700, 3.510               & 2.174 & 2.245 & 2.243      & 2.264, 2.667 & 2.334 & 2.325, 2.292       \\ 
               &Te & 4.426, 2.868 & 3.971,3.970 & 3.930, 3.928    & 2.163 & 2.228 & 2.231       & 2.501, 2.793 & 2.598, 2.596 & 2.578, 2.580  \\  \hline 
               &S & 3.199, 3.539 & 3.477 & 3.463, 3.462                 & 2.152 & 2.212 & 2.211       & 2.092, 2.068 & 2.041 & 2.038                 \\ 
MnPX$_3$  &Se & 3.983, 2.883 & 3.660 & 3.657, 3.655              & 2.171 & 2.234 & 2.232    & 2.267, 2.687 & 2.217 & 2.213                  \\ 
             &Te & 4.236, 3.162 & 3.929 & 3.969, 3.967                & 2.160 & 2.239 & 2.240      & 2.505, 2.762 & 2.487 & 2.470                  \\ \hline 
              &S & 3.307 & 3.403, 3.397 & 3.463, 3.608                            & 2.140 & 2.194 & 2.198      & 2.106 & 2.042 & 2.054, 2.047                           \\ 
FePX$_3$  &Se & 3.540 & 3.705, 3.415 & 3.623, 3.624              & 2.158 & 2.216 & 2.206      & 2.304 & 2.220, 2.230 & 2.236                 \\ 
              &Te & 3.853, 3.855 &  &                                     & 2.16 &  &                         & 2.566 &  &                                            \\ \hline
                &S & 3.327 &  &                                                & 2.224 &  &                       & 2.098 &  &                                           \\ 
CoPX$_3$  &Se & 3.571 &  &                                           & 2.254 &  &                       & 2.360 &  &                                          \\ 
                 &Te & 3.863, 3.916 &  &                                   & 2.217 &  &                       & 2.590, 2.683 &   &                                  \\ \hline
                  &S & 3.329 & 3.344 & 3.339                              & 2.199 & 2.174 & 2.184      & 2.399 & 2.052 & 2.043                            \\
NiPX$_3$  &Se & 3.540 & 3.541 & 3.544,3.543                  & 2.234 & 2.236 & 2.217      & 2.227 & 2.235 & 2.224                            \\ 
                 &Te & 3.848, 3.876 &  &                                   & 2.249 &  &                      & 2.521, 2.536 &  &                                    \\ \hline 
                  &S & 3.425, 3.499 &  &                                   & 2.215 &  &                     & 2.012, 2.084 &  &                                     \\ 
CuPX$_3$  &Se & 3.574, 3.708 &  &                                & 2.237 &  &                     & 2.188, 2.248 &  &                                    \\ 
                 &Te & 3.925, 3.709 &  &                                   & 2.243 &  &                    & 2.457, 2.462 &  &                                     \\ \hline 
                  &S & 3.445, 3.443 &  &                                   & 2.209 &  &                     & 2.040 &  &                                               \\    
ZnPX$_3$  &Se & 3.652, 3.641 &  &                                & 2.232 &  &                     & 2.214, 2.213 &  &                                    \\    
                  &Te & 3.946, 4.038 &  &                                 & 2.253 &  &                     & 2.458, 2.462 &  &                                     \\ \hline \hline 
\end{tabular}} \label{tab:2dlatticecons}
\end{center}
\end{table*}
\begin{table*}[htb!]
\caption{Total energy relative to the lowest energy magnetic configuration
among ferromagnetic (FM), antiferromagnetic (AFM) and non-magnetic (NM) states
in single layer transition metal phosphorous trichalcogenide MPX$_3$.  The absence of an entry for 
a magnetic configuration means that the corresponding state is not metastable.  Energies are  in meV/unit cell units.
Results for for two different exchange-correlation energy functional approximations are compared.  
Our first principles calculations suggest stable magnetic phases for V, Cr, Mn, Fe, Ni metal compounds.
}
\begin{center}
\begin{tabular}{cc|ccc|ccc}\\ \hline \hline
 MPX$_3$   & X &   \multicolumn{3}{c|}{GGA   (Triangular)}  &  \multicolumn{3}{c}{LDA   (Triangular)}   \\ \hline
                   &&{NM }& {FM } &{ AFM }&{NM } &{FM } &{AFM } \\ \hline
                   &S& 1133.2 & 665.4 & 0.0 & 263.8 & 276.8 & 0.0  \\
VPX$_3$    &Se& 1389.1 & 405.6 & 0.0 & 503.7 & 155.1 & 0.0  \\
                   &Te& 1237.6 &1030.0  & 0.0 & 275.3 & 724.3 & 0.0  \\ \hline
                   &S& 1220.2 & 0.0 & 41.78 & 275.7 & 0.0 & 251.3 \\
CrPX$_3$   &Se& 1737.4 & 0.0 & 331.7 & 737.5 & 0.0 & 328.6 \\
                  &Te&1479.4  & 0.0 &286.2  &881.0  &0.0  &276.9  \\ \hline
                  &S& 2104.9 & 186.9 & 0.0 & 439.7 & 0.0 & 352.2  \\
MnPX$_3$ &Se& 1351.9 & 157.7 & 0.0 & 510.6 & 0.0 & 274.0 \\
                  &Te & 792.9 &17.35  & 0.0 & 430.8 & 0.0 &66.08  \\ \hline
                  &S& 141.1& 0.0 & 329.7 & 0.0 & &   \\
FePX$_3$  &Se& 0.0 & 162.2 & 312.2 & 0.0 & &   \\
                  &Te& 0.0 &  &  & 0.0 &  &  \\ \hline
                  &S& 0.0 &  &  & 0.0 &  &  \\
CoPX$_3$ &Se& 0.0 &  &  & 0.0 &  &  \\
                 &Te& 0.0 &  &  & 0.0 &  &   \\ \hline
                 &S&435.8  & 207.48 & 0.0 & 79.55 & 44.89 & 0.0 \\
NiPX$_3$  &Se& 166.5 &152.5  &0.0  &0.0  &  &  \\
                 &Te& 0.0 &  &  & 0.0 &  &  \\ \hline
                 &S& 0.0 &  &  & 0.0 &  &  \\
CuPX$_3$ &Se& 0.0 &  &  & 0.0 &  &  \\
                 &Te& 0.0 &  &  & 0.0 &  &  \\ \hline
                 &S& 0.0 &  &  & 0.0 &  &  \\ 
ZnPX$_3$ &Se& 0.0 &  &  & 0.0 &  &  \\ 
                &Te& 0.0 &  &  & 0.0 &  &  \\ \hline \hline
\end{tabular} \label{tab:totalenergies}
\end{center}
\end{table*}

\begin{figure*}[htb!]
\begin{center}
\includegraphics[width=18cm]{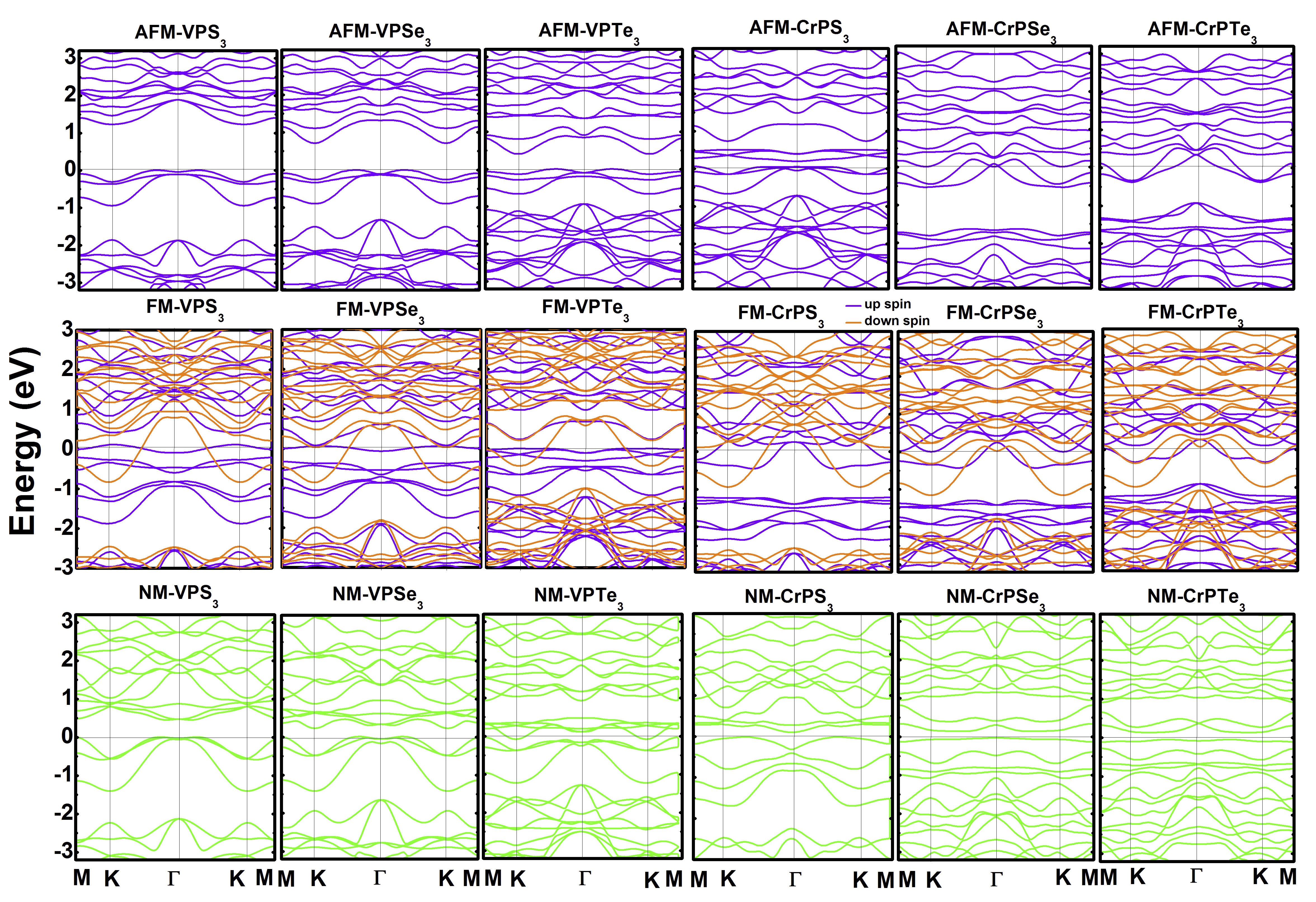}
\caption{(Color online) Band structure of single layer MPX$_3$  for M~=~V and Cr  transition metal atoms
and X~=~S, Se, Te chalcogens obtained within GGA+D2 approximation calculated for the smallest triangular unit cell.
We represent the band structure with violet for AFM, violet and orange for the  
up and down split spin bands  in the FM phase, and green colour for NM phases. 
We note that at charge neutrality the AFM phases have semiconducting band gaps, the FM phases are metallic, 
and the non-magnetic phases can either be metallic or semiconducting. 
An overall reduction of the band gaps is observed when we use heavier chalcogen atoms in keeping with the 
reduction of the covalent bond energy in larger shell orbitals. 
} \label{fig:bandstructure}
\end{center}
\end{figure*}

\begin{figure*}[htb!]
\begin{center}
\includegraphics[width=18cm]{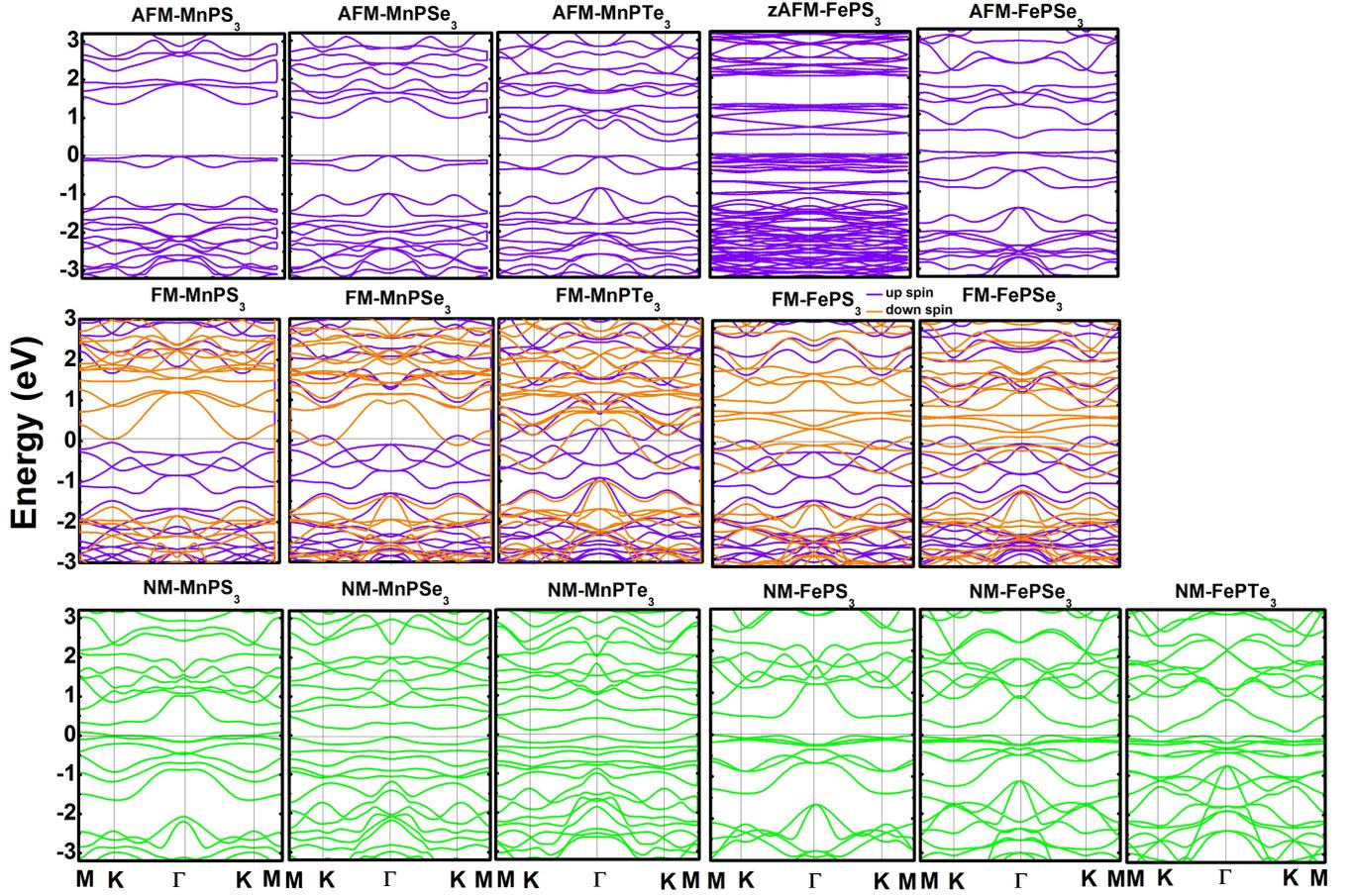}
\caption{(Color online) Band structure of single layer MPX$_3$  for M~=~Mn and Fe  transition metal atoms
and X~=~S, Se, Te chalcogens obtained within GGA+D2 approximation calculated for the smallest triangular unit cell.
We represent the band structure with violet for AFM, violet and orange for the  
up and down split spin bands  in the FM phase, and green colour for NM phases. 
We note that at charge neutrality the AFM phases have semiconducting band gaps, the FM phases are metallic, 
and the non-magnetic phases can either be metallic or semiconducting. 
An overall reduction of the band gaps is observed when we use heavier chalcogen atoms in keeping with the 
reduction of the covalent bond energy in larger shell orbitals. 
} \label{fig:bandstructure}
\end{center}
\end{figure*}

\begin{figure*}[htb!]
\begin{center}
\includegraphics[width=18cm]{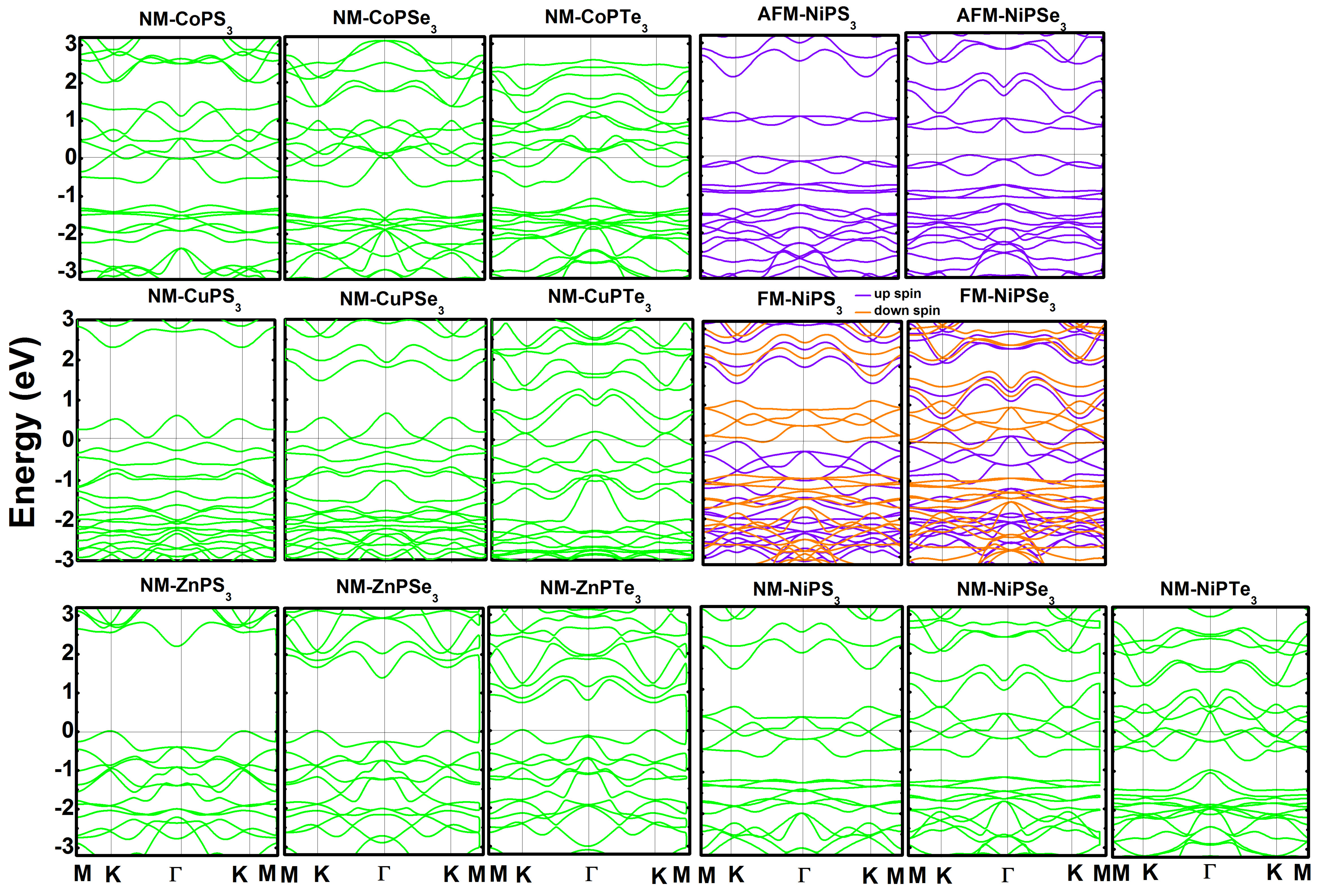}
\caption{(Color online) Band structure of single layer MPX$_3$  for M~=~Co, Ni, Cu and Zn  transition metal atoms
and X~=~S, Se, Te chalcogens obtained within GGA+D2 approximation calculated for the smallest triangular unit cell.
We represent the band structure with violet for AFM, violet and orange for the  
up and down split spin bands  in the FM phase, and green colour for NM phases. 
We note that at charge neutrality the AFM phases have semiconducting band gaps, the FM phases are metallic, 
and the non-magnetic phases can either be metallic or semiconducting. 
An overall reduction of the band gaps is observed when we use heavier chalcogen atoms in keeping with the 
reduction of the covalent bond energy in larger shell orbitals. 
} \label{fig:bandstructure}
\end{center}
\end{figure*}

\begin{figure*}[hp!]
\begin{center}
\includegraphics[width=15cm]{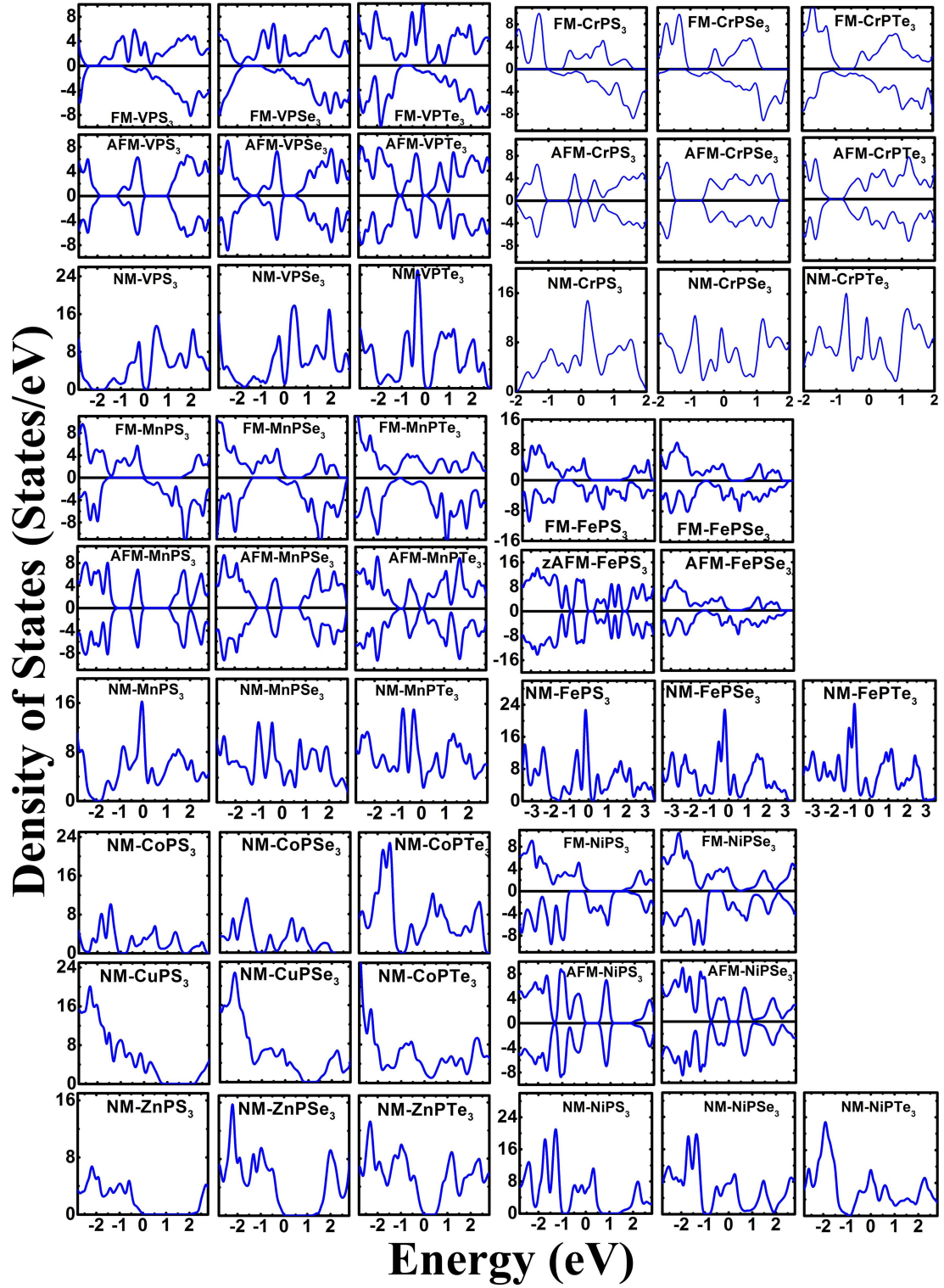}
\caption{(Color online) The density of states (DOS) calculated for every self-consistently converged magnetic configurations with total energy local minima
with the valence band edge located at $E=0$.
 } \label{fig:tdos}
\end{center}
\end{figure*}


\begin{figure*}[hp!]
\begin{center}
\includegraphics[width=15cm]{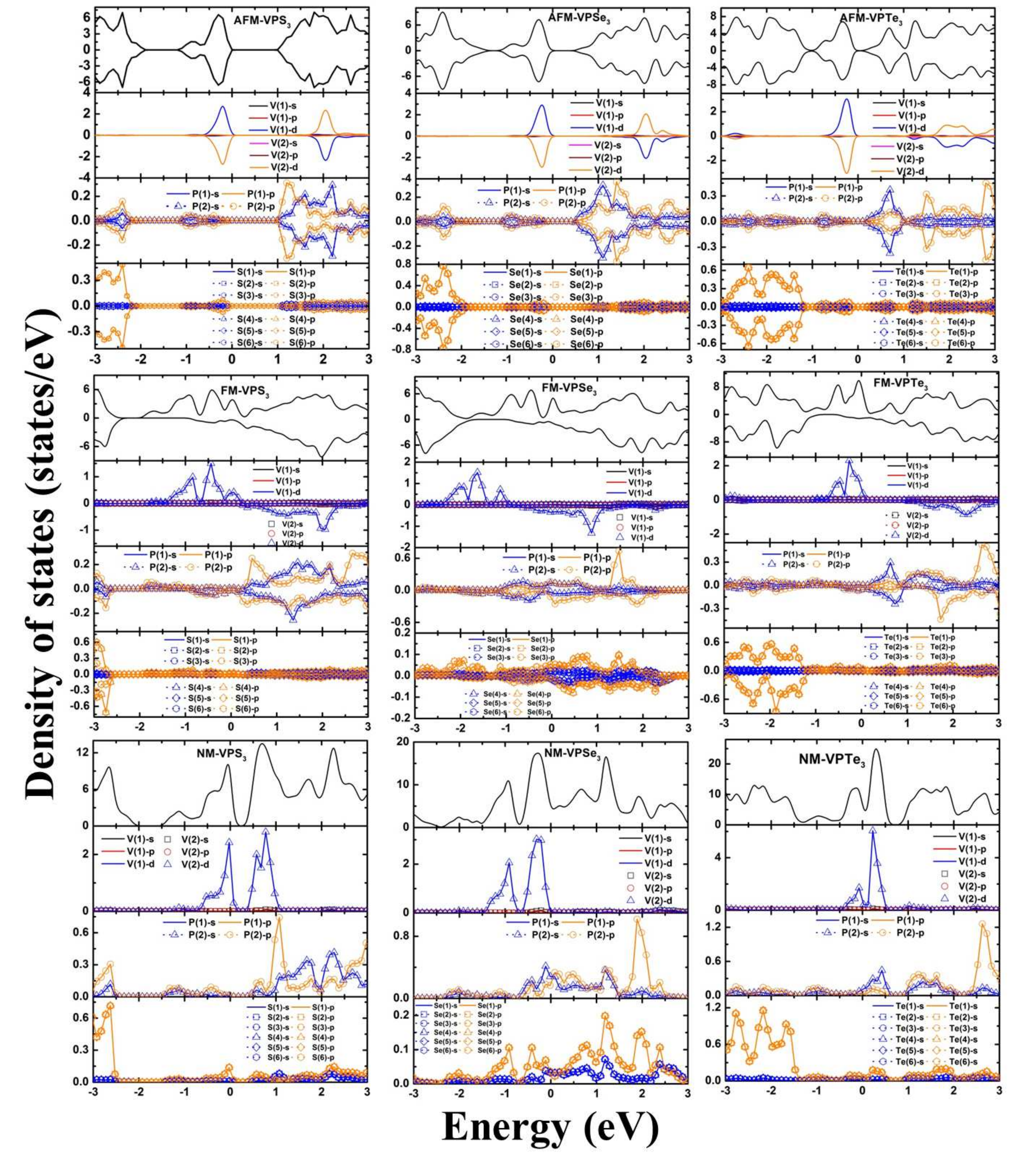} 
\caption{(Color online) Orbital projected partial density of states (PDOS) calculated for every self-consistently converged magnetic configurations with total energy local minima
with the valence band edge located at $E=0$. 
From the PDOS of the various VPX$_3$ compounds we observe that the orbital content of the valence and conduction band edges
do not generally consist of the same atomic orbital contributions and 
have a wide variation span depending on the specific choice for each one of the atoms constituting the material and the 
magnetic configuration.
 } \label{fig:vpdos}
\end{center}
\end{figure*}

\begin{figure*}[hp!]
\begin{center}
\includegraphics[width=15cm]{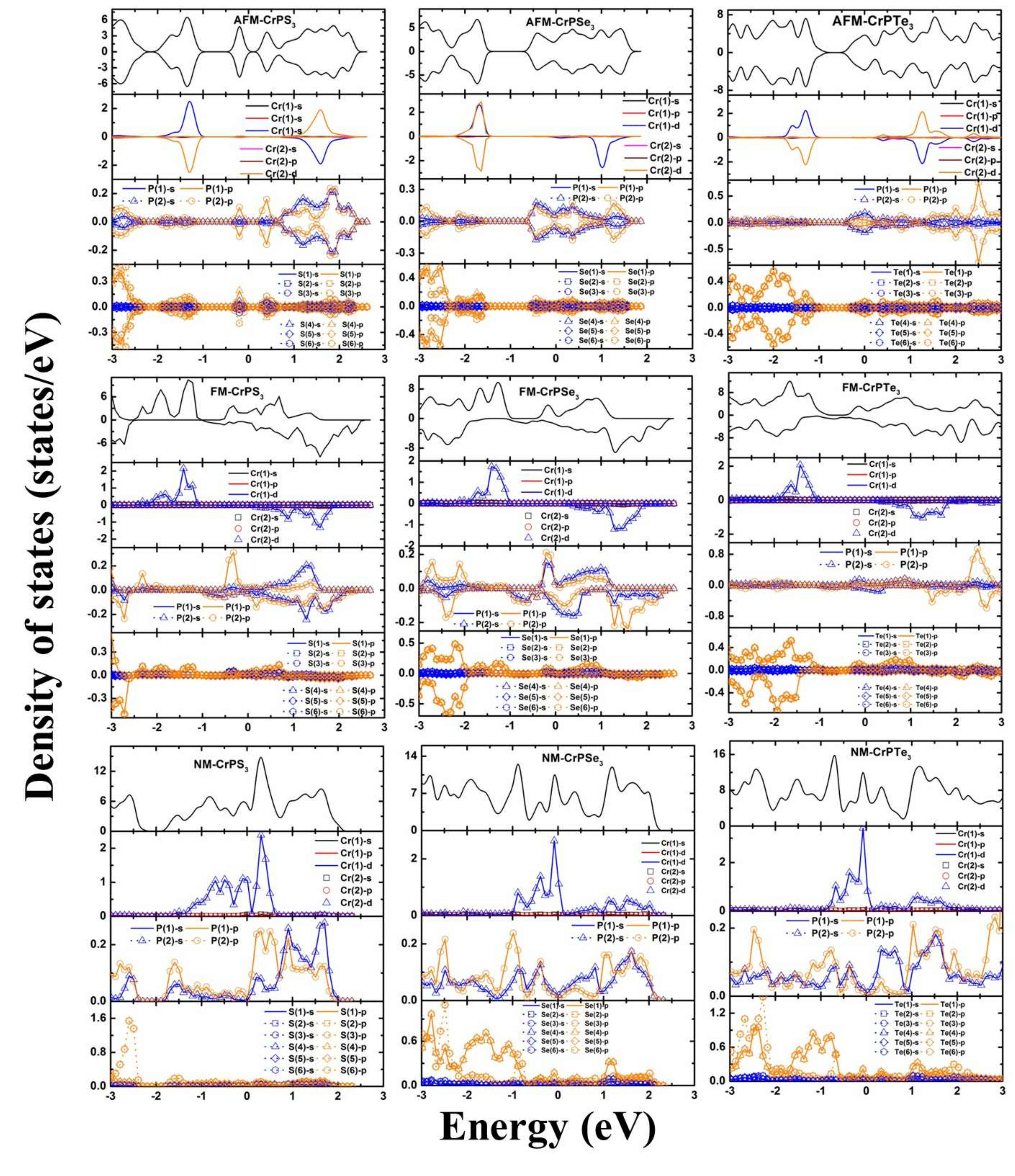} 
\caption{(Color online) Orbital projected partial density of states (PDOS) calculated for every self-consistently converged magnetic configurations with total energy local minima
with the valence band edge located at $E=0$. 
From the PDOS of the various CrPX$_3$ compounds we observe that the orbital content of the valence and conduction band edges
do not generally consist of the same atomic orbital contributions and 
have a wide variation span depending on the specific choice for each one of the atoms constituting the material and the 
magnetic configuration.
 } \label{fig:vpdos}
\end{center}
\end{figure*}

\begin{figure*}[hp!]
\begin{center}
\includegraphics[width=15cm]{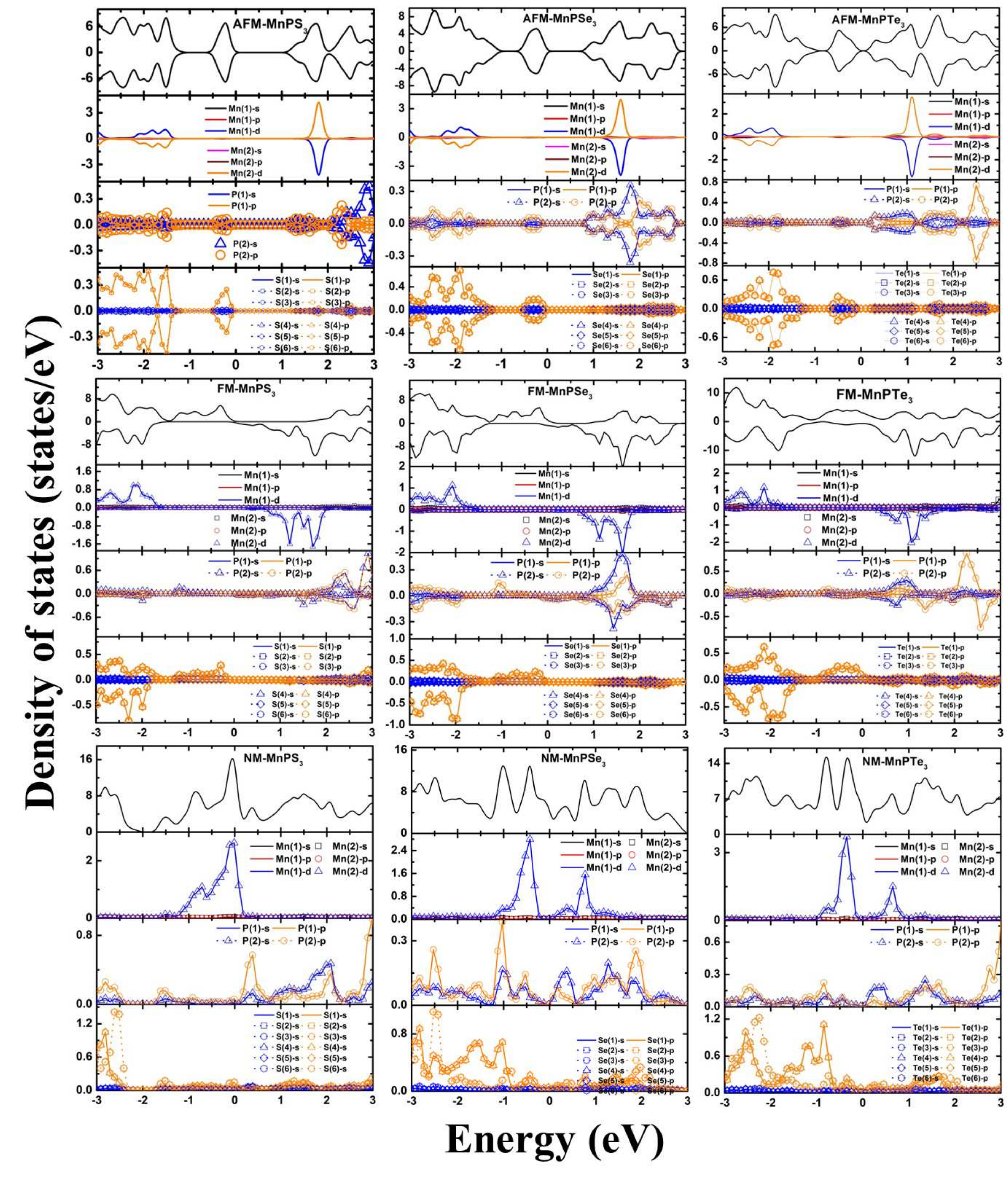} 
\caption{(Color online) Orbital projected partial density of states (PDOS) calculated for every self-consistently converged magnetic configurations with total energy local minima
with the valence band edge located at $E=0$. 
From the PDOS of the various MnPX$_3$ compounds we observe that the orbital content of the valence and conduction band edges
do not generally consist of the same atomic orbital contributions and 
have a wide variation span depending on the specific choice for each one of the atoms constituting the material and the 
magnetic configuration.
 } \label{fig:vpdos}
\end{center}
\end{figure*}

\begin{figure*}[hp!]
\begin{center}
\includegraphics[width=15cm]{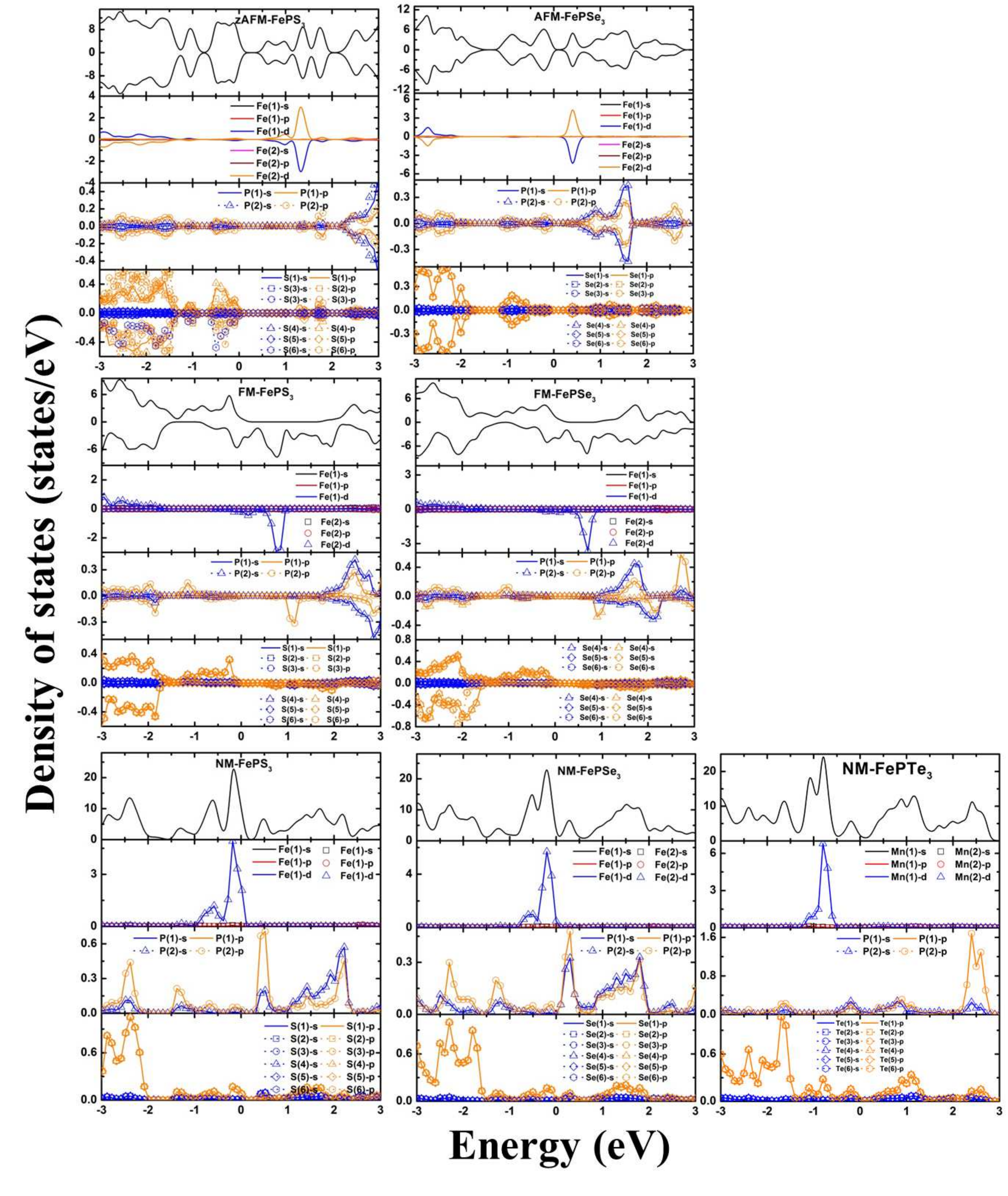} 
\caption{(Color online) Orbital projected partial density of states (PDOS) calculated for every self-consistently converged magnetic configurations with total energy local minima
with the valence band edge located at $E=0$. 
From the PDOS of the various FePX$_3$ compounds we observe that the orbital content of the valence and conduction band edges
do not generally consist of the same atomic orbital contributions and 
have a wide variation span depending on the specific choice for each one of the atoms constituting the material and the 
magnetic configuration.
 } \label{fig:vpdos}
\end{center}
\end{figure*}

\begin{figure*}[hp!]
\begin{center}
\includegraphics[width=15cm]{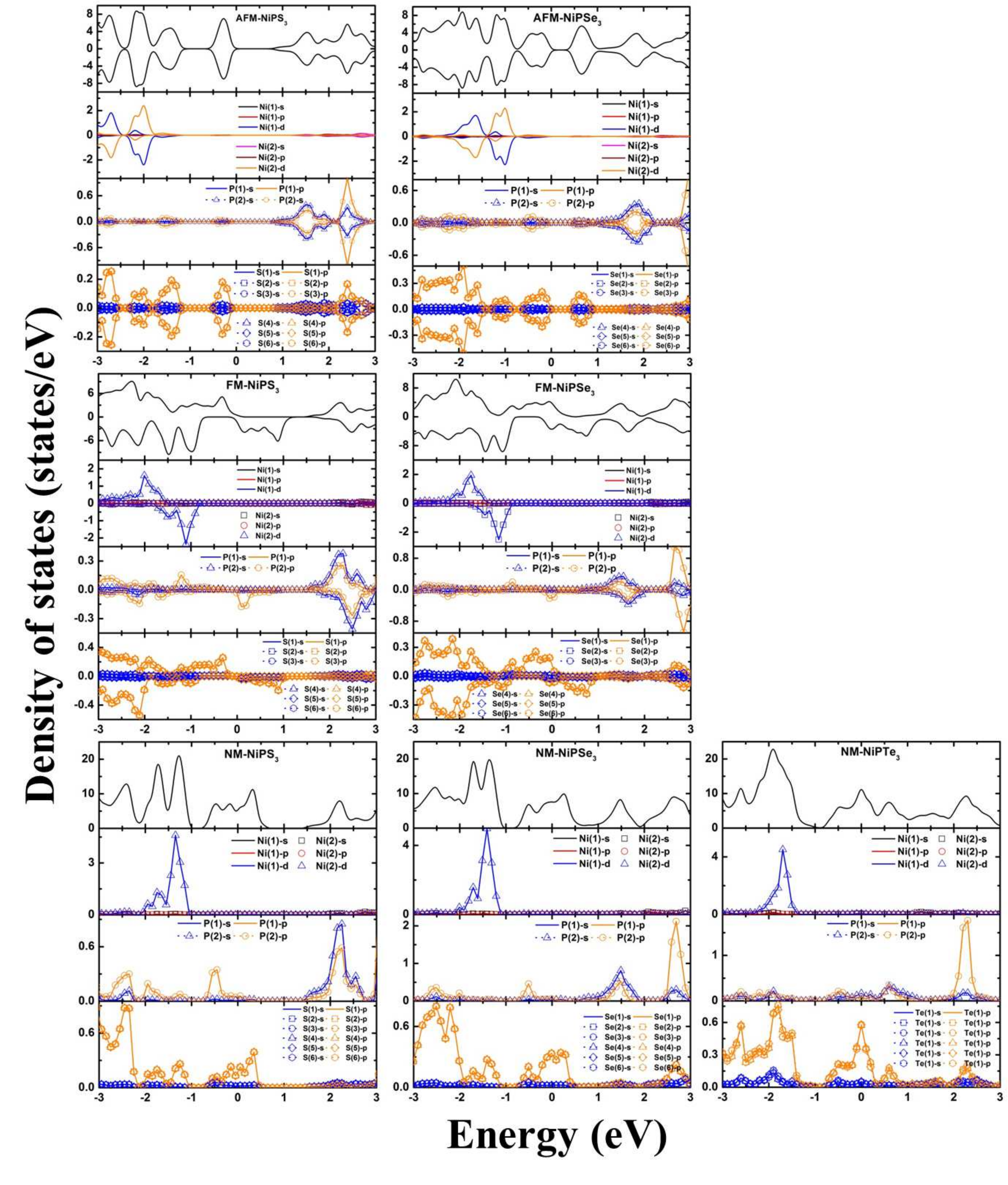} 
\caption{(Color online) Orbital projected partial density of states (PDOS) calculated for every self-consistently converged magnetic configurations with total energy local minima
with the valence band edge located at $E=0$. 
From the PDOS of the various NiPX$_3$ compounds we observe that the orbital content of the valence and conduction band edges
do not generally consist of the same atomic orbital contributions and 
have a wide variation span depending on the specific choice for each one of the atoms constituting the material and the 
magnetic configuration.
 } \label{fig:vpdos}
\end{center}
\end{figure*}

\begin{figure*}[hp!]
\begin{center}
\includegraphics[width=15cm]{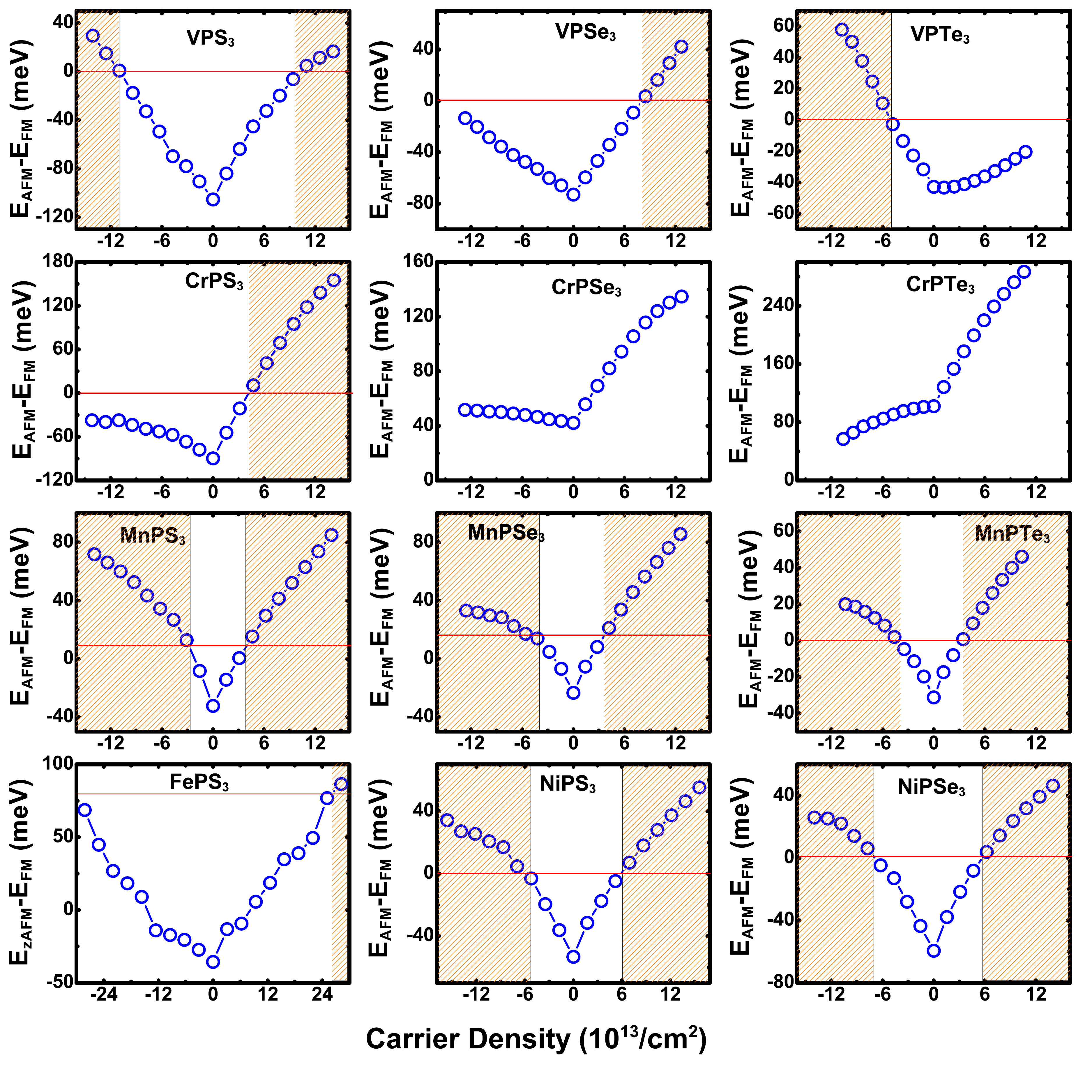}
\caption{(Color online) Carrier density dependent total energy differences per MPX$_3$ formula unit 
between the AFM and FM states of V, Cr, Mn, Fe, Ni based single layer trichalcogenides with GGA+D2+U(=4eV) calculated using a trianguar unit cell showing the modifications in the magnetic phase diagram due to the enhancement of electron-electron interactions.
 } \label{fig:magnetization}
\end{center}
\end{figure*}

\begin{figure*}[hp!]
\begin{center}
\includegraphics[width=15cm]{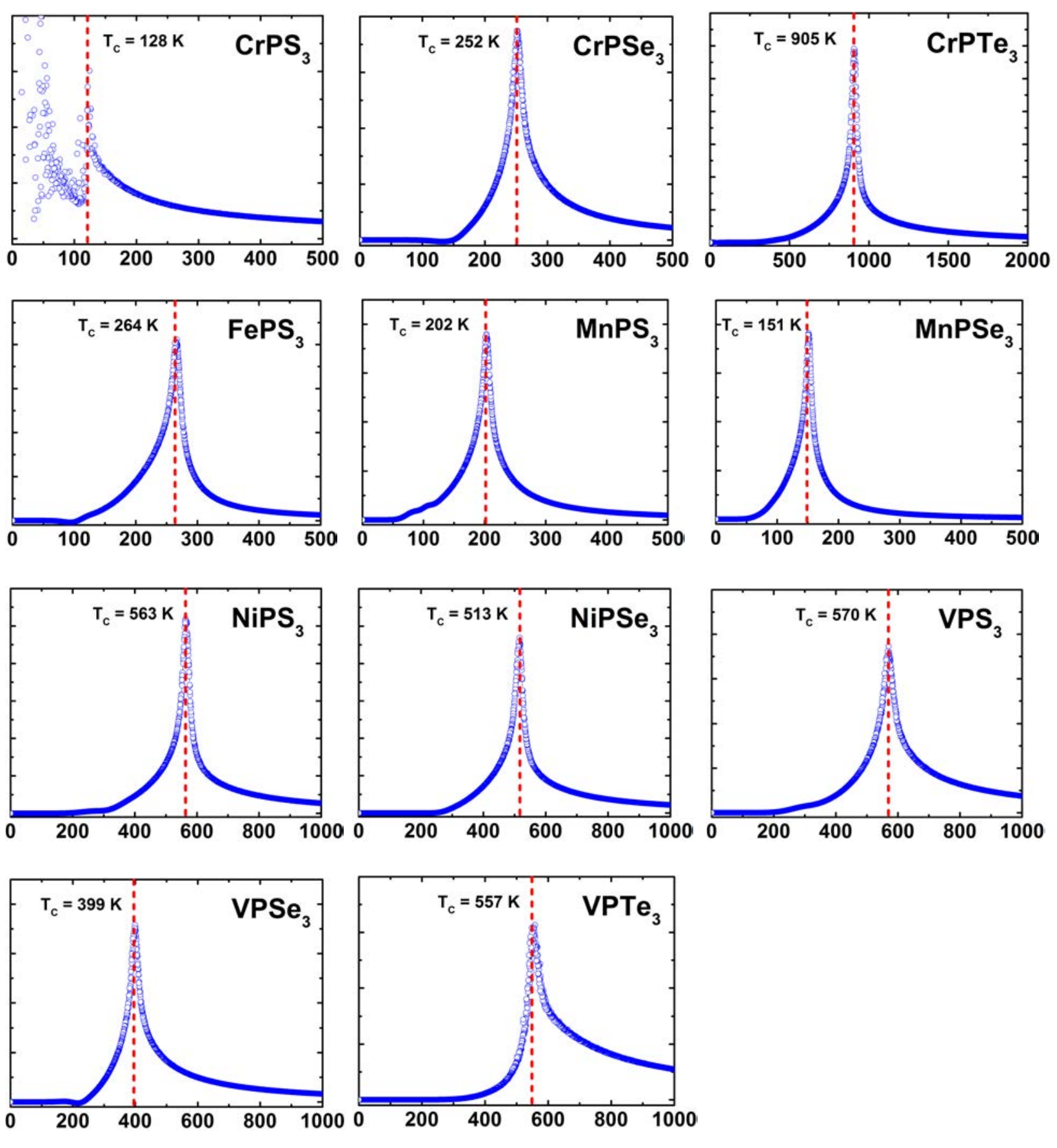}
\caption{(Color online) Temperature dependent evolution of the heat capacity 
$C= k \beta^2 \left( \langle E^2 \rangle - \langle E \rangle^2 \right)$ calculated with the Metropolis Monte Carlo algorithm in a 32$\times$64 superlattice for the effective Ising model, calculated using three nearest neighbor J-parameters obtained from the GGA+D2+U total energies where U=4eV as shown in Table III in the main text. 
The transition temperature is estimated from the maximum in the curve.
 } \label{fig:magnetization}
\end{center}
\end{figure*}

\begin{table*}[htb!]
\caption{Total energy relative to the lowest energy magnetic configuration
among ferromagnetic (FM), antiferromagnetic (AFM) and non-magnetic (NM) states
in single layer transition metal phosphorous trichalcogenide MPX$_3$ for rectangular and triangular unit cells within GGA+D2+U(=4eV). { The smaller triangular unit cell is used to compare the NM, FM and Neel AFM phases whereas the larger rectangular
unit cell is used when we also compare with the total energies of zAFM, sAFM phases.} The absence of an entry for 
a magnetic configuration means that the corresponding state is not metastable.  Energies are  in meV/unit cell units.
Our first principles calculations suggest stable magnetic phases for V, Cr, Mn, Fe, Ni metal compounds. The work functions are also calculated within triangular unitcell.
}
\begin{center}
\begin{tabular}{cc|ccccc|ccc|ccc}\\ \hline \hline
 MPX$_3$   & X &   \multicolumn{5}{c|}{Rectangular} & \multicolumn{3}{c|}{Triangular}  & \multicolumn{3}{c}{Work function (eV)}  \\ \hline
                   & & NM & FM & AFM & zAFM & sAFM &  {NM }& {FM }&{ AFM }& NM & FM & AFM \\ \hline
                   & S  & 12175 &423.4  & 0.0 &251.9 &130.6  & 210.6   & 213.6 & 0.0 &1.41  &2.58  &3.12  \\
VPX$_3$    & Se& 10055 &298.2  & 0.0 &169.9 &93.2  & 149.0  & 148.4 &0.0  &  1.73&2.36  &2.05  \\
                  & Te& 9564 & 183.1 & 0.0 &648.9 & 48.9 &4850    & 85.6 & 0.0 &  1.20& 1.59 &1.31  \\ \hline
                  & S  & 10282 &163.2  &0.0  &188.3 & 8.16 &  5553  &179.3  &0.0  &1.67  &2.17  &2.36  \\
CrPX$_3$  & Se& 10583 &60.6  &0.0  & 148.7& 47.3 &  5587  &0.0  & 84.1 & 1.67 &1.90  &2.04  \\
                  & Te& 10855 &213.8  &478.7  &504.4 &0.0  & 5877   &0.0  &204.2  & 0.91 &1.23  &1.49  \\ \hline
                  & S  &12807  & 129.2 & 0.0 & 51.7&  58.7&  7013  & 64.6 &0.0  &2.05  &1.53  &1.77  \\
MnPX$_3$ & Se&15744  &92.5  &0.0  &31.2 &47.2  & 6578   & 46.6 & 0.0& 1.94 &1.60  &1.71  \\ \hline
FePX$_3$  & S&4779  &183.0  &0.0  &28.5 &100.6  &2433    &0.0  & 1327 &3.01  &2.67  &2.71  \\ \hline
NiPX$_3$   & S &2621  &318.6  &  91.1&0.0  &321.1 &1292  & 106.6   &0.0  &1.99  &2.26  &2.00    \\
                 & Se & 2036 &292.5  &54.4  & 0.0& 303.4 &  996.7  & 118.9 & 0.0 & 1.48 & 1.81 &2.14  \\ \hline \hline
\end{tabular} \label{tab:totalenergies}
\end{center}
\end{table*}

\end{document}